\documentclass[a4paper, table]{article}

\usepackage[utf8]{inputenc} 
\usepackage{amsmath, amsthm, amssymb, amsfonts} 
\usepackage{lipsum} 

\usepackage{graphicx} 
\usepackage{float} 
\usepackage{caption} 
\usepackage{subcaption} 

\usepackage{tabularx} 
\usepackage{tabu} 
\usepackage{booktabs} 

\usepackage[nottoc,numbib]{tocbibind} 
\usepackage[margin = 2.5cm]{geometry} 
\usepackage{microtype} 
\usepackage{titlesec} 
\usepackage{titletoc} 
\usepackage{appendix} 
\usepackage{fancyhdr} 
\usepackage[shortlabels]{enumitem} 

\usepackage{hyperref} 
\usepackage[noabbrev, capitalise]{cleveref} 

\usepackage{xcolor} 

\usepackage{tikz} 
\usetikzlibrary{positioning} 
\usetikzlibrary{calc} 

\makeatletter
\newcommand{\gettikzxy}[3]{%
  \tikz@scan@one@point\pgfutil@firstofone#1\relax
  \edef#2{\the\pgf@x}%
  \edef#3{\the\pgf@y}%
}
\makeatother

\newcommand\reporttitle{An Educational Guide for 2D Stellar \\ \bigskip Structure Calculations of Rapidly \\ \bigskip Rotating Stars using the  ESTER code}

\newcommand\reportsubtitle{ 
}

\newcommand\groupnumber{
\textbf{}
}

\usepackage{orcidlink}
\newcommand\reportauthors{
Russel White$^{b,a}$  \orcidlink{0000-0001-5313-7498} (white@chara.gsu.edu) \\
 Jane Pratt$^{a,b}$  \orcidlink{0000-0003-2707-3616} (pratt34@llnl.gov) \\
 Michel Rieutord$^{c}$ \orcidlink{0000-0002-9395-6954} (mrieutord@irap.omp.eu)
}

\newcommand\grouptutor{
\begin{enumerate}[a)]
\item Astronomy and Astrophysics Analytics Group, Lawrence Livermore National Laboratory, 7000 East Ave, Livermore, CA 94550, USA 
\item Department of Physics and Astronomy, Georgia State University, Atlanta GA 30303, USA 
\item IRAP, Université de Toulouse, CNRS, CNES, 14, avenue Édouard Belin, F-31400 Toulouse, France
\end{enumerate}
}

\newcommand\placeanddate{
Livermore, California \today
}

\definecolor{Tue-red}{RGB}{199, 25, 24}
\definecolor{lightblue}{rgb}{.8, 1., 1.}
\definecolor{cadet}{rgb}{.3725, .619, .627}
\definecolor{cyan}{rgb}{0., .545, .545}
\definecolor{sea}{rgb}{.235, .702, .443}
\definecolor{aqua}{rgb}{.561, .737, .561}
\definecolor{turq}{rgb}{.686, .9333, .9333}
\definecolor{whiteblue}{rgb}{0.2, .8, .6}
\definecolor{bluey}{rgb}{0.2, .8, 1.}
\definecolor{ltblue4}{rgb}{.902, .957, 1}
\definecolor{seablue}{rgb}{.3725,  .619,  .627}
\definecolor{dodger}{rgb}{.0,.2758,.5151}

\hypersetup{
    colorlinks=true,
    linkcolor=dodger,
    urlcolor=dodger,
    citecolor = dodger
    }
\counterwithin{equation}{section} 
\counterwithin{figure}{section} 
\counterwithin{table}{section} 

\titleformat{\section}{\sffamily\color{dodger}\Large\bfseries}{\thesection\enskip\color{gray}\textbar\enskip}{0cm}{} 

\titleformat{\subsection}{\sffamily\color{dodger}\large\bfseries}{\thesubsection\enskip\color{gray}\textbar\enskip}{0cm}{} 

\titleformat{\subsubsection}{\sffamily\color{dodger}\bfseries}{\thesubsubsection\enskip\color{gray}\textbar\enskip}{0cm}{} 

\DeclareCaptionFont{dodger}{\color{dodger}}
\usepackage{mlmodern}

\usepackage{graphicx}
\usepackage{xspace}
\usepackage{amsmath}
\usepackage{amsfonts}
\usepackage{amssymb}
\usepackage{placeins}
\usepackage{gensymb}
\usepackage{color}
\usepackage{float}
\usepackage{hyperref}
\usepackage[english]{babel}
\usepackage{natbib}
\usepackage{ragged2e} 
\usepackage[skip=10pt plus 10pt, indent=10pt]{parskip}

\renewcommand{\vec}[1]{\mbox{\boldmath$#1$}}

\begin{document}

\begin{titlepage}

\centering

\begin{tikzpicture}

\node[opacity=0.2,inner sep=0pt,remember picture,overlay] at (4.5,-0.5){\includegraphics[width= 0.8 \textwidth]{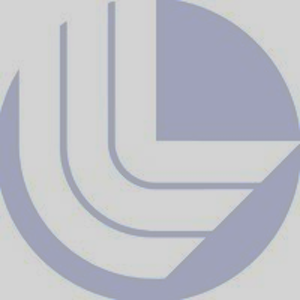}};

\node[inner sep=0pt] (logo) at (0,0)
    {\includegraphics[width=.25\textwidth]{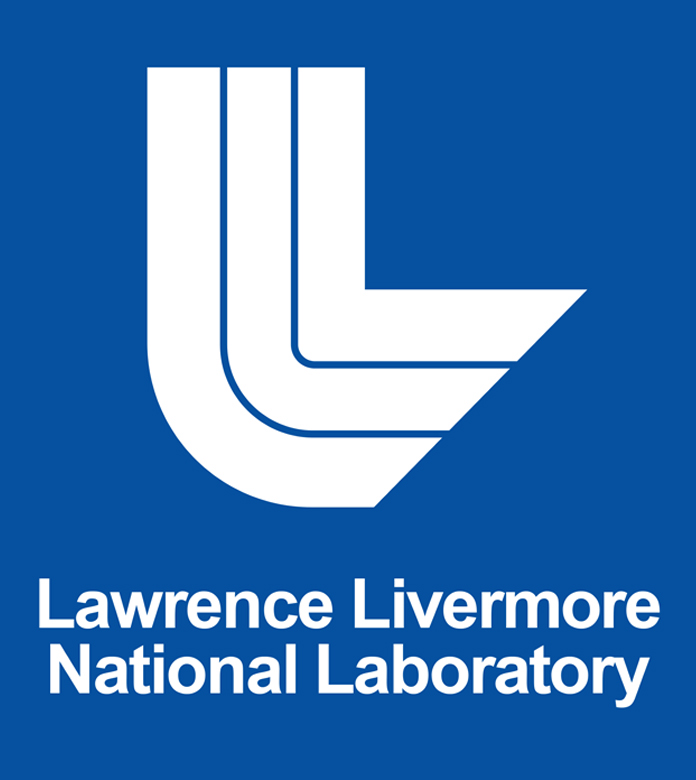}};
    
\node[text width = 0.7\textwidth, right = of logo](title){\sffamily\huge\reporttitle};

\node[text width = 0.5\textwidth, yshift = 0.75cm, below = of title](subtitle){\sffamily\Large \reportsubtitle};

\gettikzxy{(subtitle.south)}{\sffamily\subtitlex}{\subtitley}
\gettikzxy{(title.north)}{\titlex}{\titley}
\draw[line width=1mm, dodger]($(logo.east)!0.5!(title.west)$) +(0,\subtitley) -- +(0,\titley);

\end{tikzpicture}
\vspace{3cm}

\sffamily\groupnumber

\sffamily
\large

{\sffamily\reportauthors}

\sffamily \grouptutor

\tikz[remember picture,overlay]\node[anchor=south,inner sep=0pt] at (current page.south) {\includegraphics[width=\paperwidth]{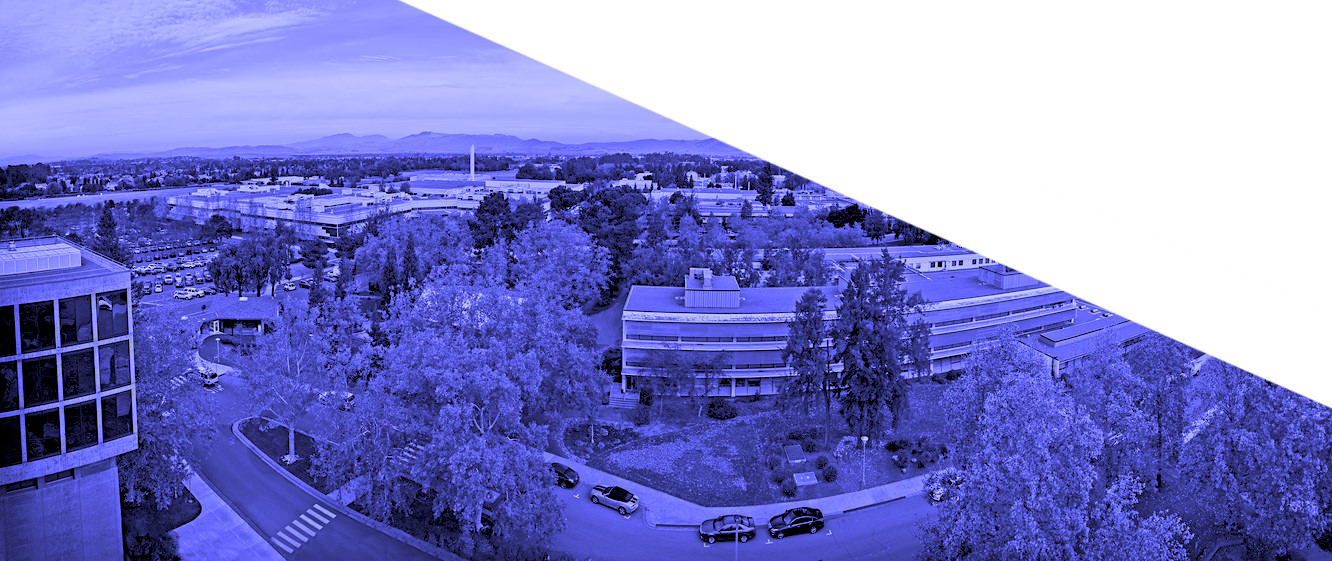}};

\mbox{}
\vfill
\sffamily \Large \textcolor{white}{\placeanddate} \\

\end{titlepage}

\newpage

{\hypersetup{linkcolor=black} 

\section*{Abstract}
\justify{ 
  The Evolution STEllaire en Rotation (ESTER) code is the first 2D stellar structure code to be made open-source and freely available to the astronomy and astrophysics community.  An important and
  novel advancement of this code is that it can reproduce the distorted shape and observable signatures (e.g., gravity darkening) of rapidly rotating stars. {ESTER also calculates the steady-state large-scale flows within the star, namely their differential rotation and associated meridional circulation}.  In this report, we explore and document the physics implemented within version 1.1.0rc2 of the ESTER code, in a way that complements published descriptions.  We illustrate this physics by plotting how stellar structure parameters vary through stellar interiors at a range of latitudes and at different angular velocities.  We investigate how the thin convective envelopes of intermediate mass stars vary with latitude when rapidly rotating, becoming deeper and thicker near the equator.  Simple comparisons of ESTER model predictions (e.g., central temperature and density, luminosity) with the output from the Modules for Experiments in Stellar Astrophysics (MESA) code \citep{paxton2010modules} shows generally good agreement. Additional comparisons provide important benchmarking and verification for ESTER as a comparatively young code. Finally, we provide a guide for installing and running the code on our local university cluster, aimed at helping students to begin work.}

\vspace*{2mm}
\tableofcontents\thispagestyle{empty}}

 \pagenumbering{arabic}

\setlength{\parskip}{16pt}

\section{\label{sec:intro} Introduction}

\subsection{Stellar rotation \label{sec:rotation}}

Angular momentum is a fundamental property of a star. It can alter mixing throughout the stellar interior, change the shape of a star, and shift the course of its evolution. Observational evidence suggests that stars are born rotating rapidly \citep[e.g.,][]{wolff_simon1997}, but the rotational evolution of stars is strongly mass dependent \citep{maeder2000evolution}. Low mass stars spin-down comparatively quickly \citep{matt2015mass} while intermediate and higher mass stars spin-down more slowly. This division occurs within the mass range of 1.3 - $1.4\,M_{\odot}$, corresponding roughly to spectral type F5 \citep{beyer_white2024}. For stars less massive than 1.3 - $1.4\,M_{\odot}$, the outer convective envelope generates a magnetic field that couples to the stellar wind; the Sun is a convenient example of this. A large-scale magnetic field has the effect of slowing down a star's rotation rate over time through the mechanism of \emph{magnetic braking} \citep[e.g.,][]{schatzman1962, weberdavis1967, skumanich1972, matt2010spin, denissenkov2010model, meynet2011massive, matt2012, gallet_bouvier2013, pantolmos2017magnetic, vansaders2019,takahashi2021modeling, ireland2022magnetic, metcalfe2022weakened, lu2024empirical}. In contrast, higher mass stars, while on the main sequence, have an outer radiative zone \citep[e.g.,][]{jermyn2022}.
A simplified interpretation is that these stars do not generate an external magnetic field of sufficient
strength to slow the star's rotation via magnetic braking. These stars retain the angular momentum imparted during
their pre-main sequence phase, resulting in rapid rotation.  We note however that these same intermediate and higher mass stars have convective cores \citep[e.g.,][]{jermyn2022} and MHD simulations demonstrate that rotating cores can produce large-scale magnetic fields \citep[e.g.,][]{brun2005simulations} even though the detection of these fields may be difficult.  Indeed, magnetic fields have been detected in some intermediate mass stars \citep[e.g., Vega, Sirius;][]{cantiello_braithwaite2019}, although these are often interpreted as fossil fields, i.e. a stable, or slowly decaying, magnetic field left over from an earlier epoch \citep[e.g.,][]{villebrun2019magnetic,alecian2008magnetism,featherstone2009effects}. While magnetic fields generated in intermediate and higher mass stars that have convective cores are the subject of open research, the interpretation of the effect of such magnetic fields on rotation is grounded empirically; the transition in average rotation rate is sharp and well defined.  This transition has been known for a half-century and is referred to as the \emph{Kraft break} \citep{kraft1967, jayasinghe2020asas, avallone2022rotation}.  \citet{beyer_white2024}
recently refined the location of the Kraft Break in mass, effective temperature, and optical color.

\subsection{Oblate shapes and gravity darkening \label{sec:rotationeffects}}

In many cases the rotation rate of stars above the Kraft Break (i.e., hotter than the Kraft Break) is rapid enough to distort the star into an oblate shape. It is common to quantify how much the star is ``flattened'' from this rotation  as $1 - R_{\mathsf{p}}/R_{\mathsf{e}}$ \citep[e.g.,][]{espinosa_lara2011gravity}. Here $R_{\mathsf{p}}$ is the polar radius and $R_{\mathsf{e}}$ is the equatorial radius. Using the simple Roche model \citep{vakili2002modelling,zahn2010shape}, at large rotational velocities, $R_{\mathsf{e}}$ can become as large as 1.5 $R_{\mathsf{p}}$ \citep{vanbelle2012}, corresponding to a flatness of 0.333.

In the study of rapidly rotating stars, rotation is often defined relative to a star's critical velocity, a velocity where the centripetal force equals the gravitational force, $F_{\mathsf{gravity}} = F_{\mathsf{centripetal}}$.  For a star of mass, $M_\star$, the critical velocity is defined as \citep{townsend2004star}
\begin{eqnarray}
v_{\mathsf{crit}} = \sqrt{G M_\star/R_{\mathsf{e}}} = \sqrt{2 G M_\star/ 3 R_{\mathsf{p}}}~.
\end{eqnarray}
In the present analysis, we characterize rotation by a star's \textit{angular velocity}, $\omega$, relative to its critical angular velocity, $\omega_{\mathsf{crit}}$ \citep{granada2013populations},
\begin{eqnarray}\label{transformtoangularmomentum}
\omega/\omega_{\mathsf{crit}} = \frac{v}{v_{\mathsf{crit}}}  \frac{R_{\mathsf{e,crit}}}{R_{\mathsf{e}}}~.
\end{eqnarray}
In this definition it is important to note that $R_{\mathsf{e}}$ depends on $\omega$.  The use of the angular velocity rather than the velocity is common in stellar structure and evolution modeling.

\begin{figure}[h]
\begin{center}
\resizebox{6in}{!}{\includegraphics{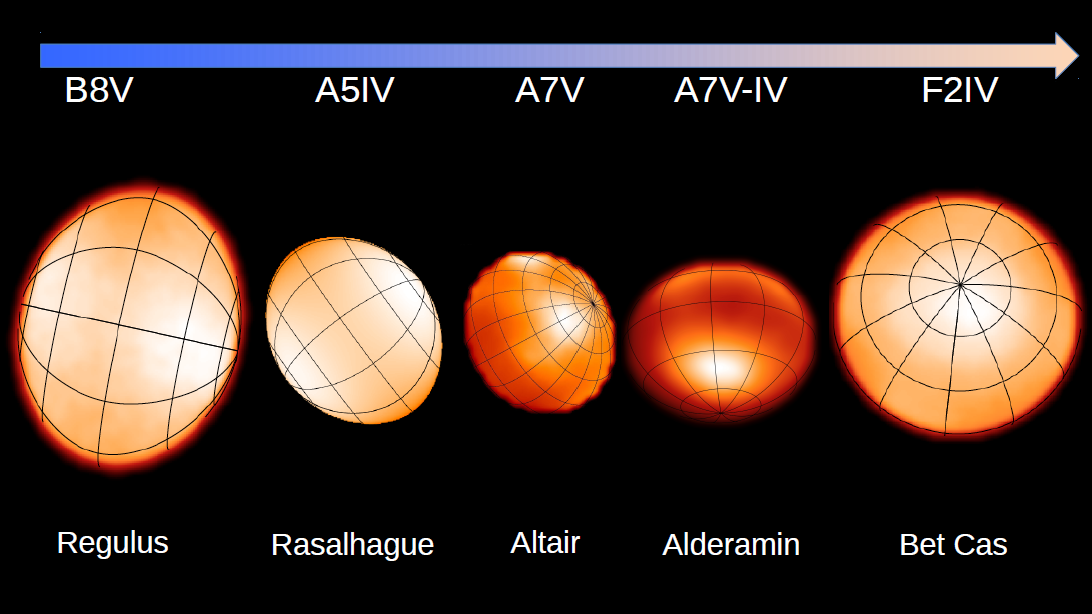}}
\caption{Images of rapidly rotating stars obtained using the CHARA Array, an optical/infrared interferometer.  Although these stars are observed at a variety of inclinations, they are all oblate in shape and exhibit gravity darkening: their equatorial regions are cooler than their polar regions.
\label{fig:rapid_rotators} }
\end{center}
\end{figure}

The oblate shape of a rapidly rotating star will also cause a phenomenon known as \emph{gravity darkening} (see Figure \ref{fig:rapid_rotators}).  Gravity darkening was predicted a century ago by \citet{vonzeipel1924}.  This work predicted that the local effective temperature, $T_{\mathsf{eff}}$, should scale with the local surface gravity, $g_{\mathsf{eff}}$, according to
\begin{eqnarray}\label{eq:vonzeipel}
{T}_{\mathsf{eff}}\propto {g_{\mathsf{eff}}}^{\beta}~,
\end{eqnarray}
where $\beta$ is referred to as the \textit{gravity darkening exponent}. This relation with a value of $\beta = 0.25$ is commonly called the von Zeipel law and is considered appropriate for stars with radiative envelopes. \citet{lucy1967} proposed a shallower dependence on local gravity, $\beta = 0.08$, for stars with convective envelopes. These power-law scalings for gravity darkening may also depend on the atmospheric model adopted in a stellar structure code, as discussed by \citet{espinosa_lara_rieutord2013}.

The distinctive signature of gravity darkening has been confirmed in images of rapidly rotating stars observed using long-baseline optical/infrared interferometers \citep[e.g.,][]{mcalister2005,ten_brummelaar2005}. These images provide direct constraints on the oblate shape as well as the magnitude of gravity darkening \citep{monnier2007, ming2009, che2011, vanbelle2012}. They demonstrate that a single value for the effective temperature or surface gravity no longer accurately describe stars that experience gravity darkening, and their observed properties are inclination dependent. Gravity darkening can also affect spectroscopic measurements of rapidly rotating stars \citep{lazzarotto+23,montesinos2024}.  If unaccounted for, this effect can skew photometric and spectroscopic observations, and thus bias observational studies of intermediate and high mass stars \citep[e.g.,][]{jones2015ursa_majoris, brandt_huang2015}.

\begin{figure}
  \begin{minipage}[c]{0.5\textwidth}
  \resizebox{3.2in}{!}{\includegraphics{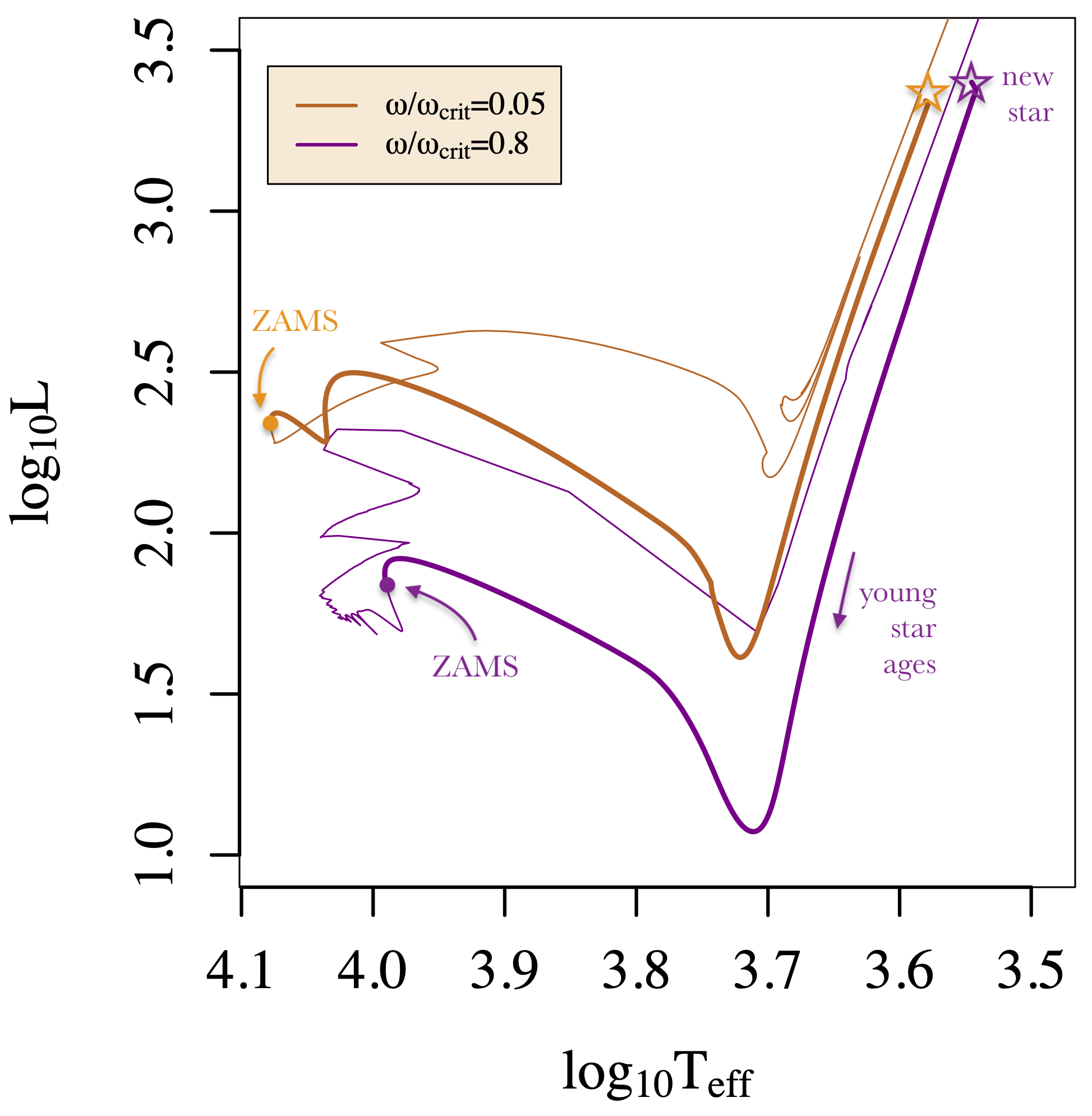}}
  \end{minipage}
  \hspace{5mm}\begin{minipage}[c]{0.4\textwidth}
      \caption{Hertzsprung--Russell (HR) diagram showing tracks of a 4 $M_{\odot}$ star with a fixed rotation rate, calculated using the MESA code.  Two stellar tracks are shown, one rotating at 5\% of the critical angular velocity and one rotating at 80\% of the critical angular velocity (rapid rotation).  The initial star is marked with a star icon, and the direction on the HR diagram that the star ages is indicated by an arrow.  The pre-main-sequence of each star is indicated by a thicker line.  A comparison of the zero age main sequence (ZAMS) positions demonstrates that rapidly rotating stars are distinctively less luminous and cooler.  These rotation rates are in contrast to the rotation rate of the Sun, which is less that 1\% of the critical angular velocity \citep{breton2022stochastic} \label{fig:mesa_resultrot}}
  \end{minipage}
\end{figure}

Stellar rotation can also affect stars in ways that are more difficult to observe.  Stellar rotation may lead to differential rotation in the stellar interior; the plasma at different internal radii of the star can rotate at different rates.  Likewise, different latitudes of a rotating star may rotate at different rates, producing zonal flows \citep{busse1983model}. The details of how rotation interacts with the other fluid effects of stellar plasmas can change the course of stellar evolution \citep{van2018sensitivity,lovekin2009effects}. Modern stellar evolutionary models that incorporate rotation predict that rapidly rotating stars will be cooler, less luminous and will evolve more slowly than slowly rotating stars.  The evolution of two intermediate mass stars with different rates of rotation can be very different (see Figure \ref{fig:mesa_resultrot}).

\subsection{ESTER: A tool for modeling rapidly rotating stars}

Progress in modeling the distorted structure of rapidly rotating stars requires a two-dimensional (2D) formalism; a 3D formalism for stellar structures would be necessary for including other physical effects like non-radial pulsations and tidal distortion \citep{lovekin2020challenges}.  The ESTER code (Evolution STEllaire en Rotation), described in \citet{rieutord2016}, is a 2D stellar structure code that has already enabled useful constraints on the effects of rapid rotation in stars, including differential rotation, meridional circulation \citep{wood2012transport, featherstone2015meridional}, and gravity darkening.  \citet{bouchaud2020} describe the development and the state of the open-source ESTER code v1.1.0rc2:
\begin{quotation}
\noindent
``Recent progress in programming techniques and computer power has enabled the creation of fully 2D stellar models by the ESTER code (Rieutord et al. 2016). ESTER models indeed predict the differential rotation profile and the associated meridional circulation of an early-type star at a given stage of its [main sequence (MS)] evolution. The solution given by the code is presently the steady state solution of an isolated rotating star. Time evolution has not been implemented yet. However, by tuning the hydrogen mass fraction in the convective core, a good approximation of an evolved state on the MS can be computed.''
\end{quotation}
The most recently released version of the code is able to compute the evolution of a rapidly rotating star for a short time along the its main-sequence evolution \citep{mombarg2023first}. Both versions of the ESTER code are open-source and available to the community via GitHub \citep{esterusersguide}; pragmatically, it is thus an available tool for the community to adopt for wide use and further development. {A technical description of the ESTER project is available on the GitHub server at \url{https://ester-project.github.io/ester/download/ester.pdf} .}

\subsection{Capabilities of ESTER v1.1.0rc2 \label{sec:capab}}

The version of ESTER that we treat here, v1.1.0rc2, is purely a stellar structure code that does not evolve the star. We focus on this version of the code to investigate, validate, and help disseminate the 2D formalism of this novel prescription. It is our hope that this will aid the code's further development.

\begin{figure}[h]
\begin{center}
\resizebox{6in}{!}{\includegraphics{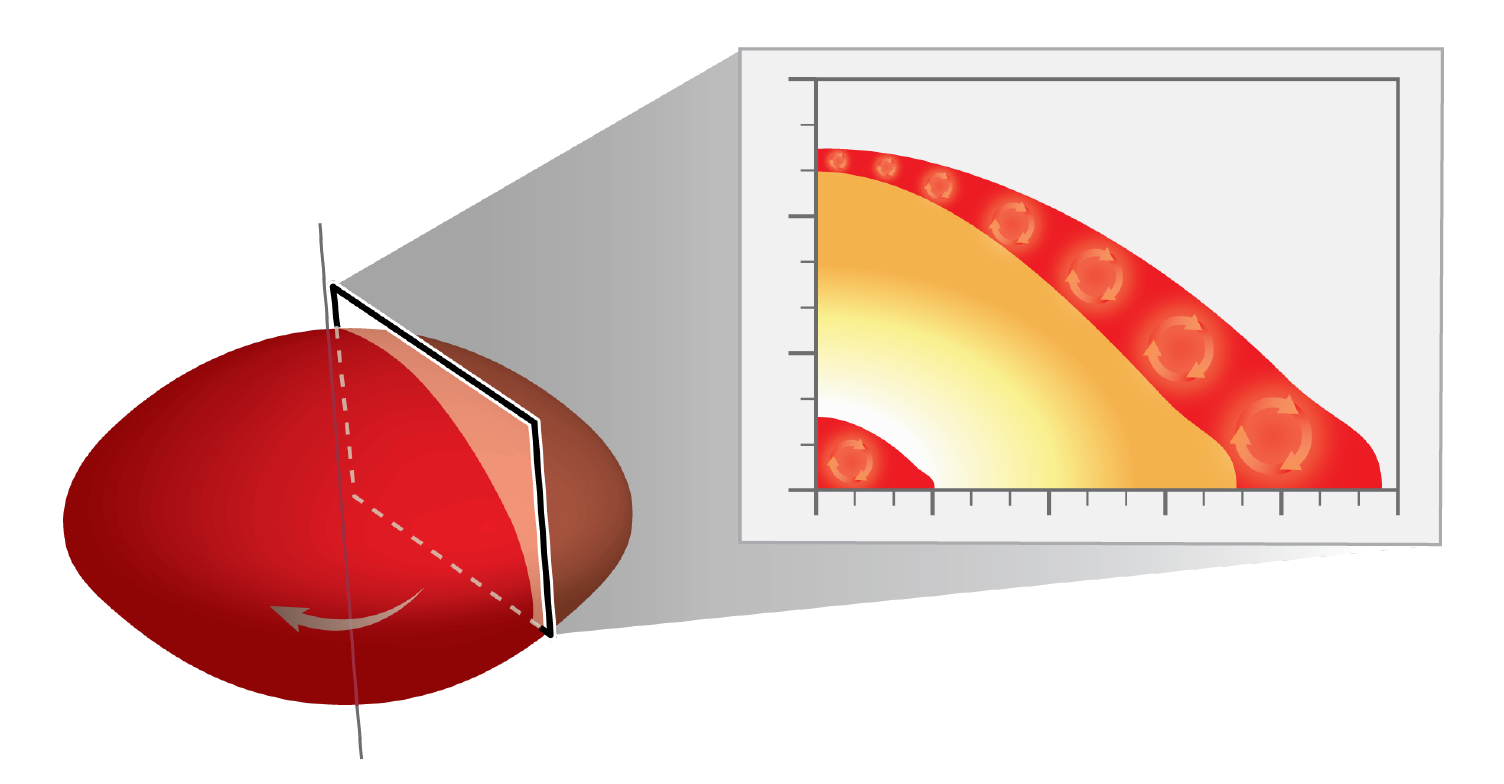}}
\caption{Conceptual sketch of a rapid rotator that has two convection
zones: a small convective core, as well as a thin convective envelope  (red areas).  These convection zones have a size
dependent on latitude.  The shaded area between them indicates the temperature decreasing from the core outward.  The
sketch includes the whole star below the photosphere.
\label{fig:fatconvzone} }
\end{center}
\end{figure}

A limitation of both versions of the ESTER code is that they do not include any prescription for convection in an outer envelope. This makes the code less suitable for modeling stars cooler than the Kraft Break; the Sun, as an example, has an outer convective envelope consisting of $\sim 2$\% of its mass, but nearly 30\% of its radius. ESTER is currently most appropriate for main sequence stars more massive than about $1.4\,M_\odot$. \citet{rieutord2013present} suggest a limiting mass that is higher than this, although they may simply be conservative in their estimate:
\begin{quotation}
\noindent ``No time evolution is included and only early-type stars (mass larger than $1.7\,M_\odot$) can be computed. The modeling of an outer convection zone is still a problem that needs to be solved. The central convective core is assumed to be isentropic.''
\end{quotation}
The general expectation is that most stars cooler than the Kraft Break are slowly rotating because they have experienced magnetic braking. The effects of rotation are likely to be less significant in these cases.  But there are some important exceptions. Young stars, intermediate mass stars transitioning across the Hertszprung Gap, and tidally interacting binary stars can have cool convective envelopes and rapid rotation. Without a prescription for an outer convective zone, ESTER cannot accurately model these stars. As 2D and 3D convection prescriptions continue to advance, ESTER may be further developed to provide a framework for modeling such stars. The illustration in Figure \ref{fig:fatconvzone} shows what a rapidly rotating star with a convective core and a convective envelope may look like. The lower density envelope, in particular, should have a convective zone depth that varies latitudinally. \citet{vanbelle2012} and others have discussed this possibility.
Since gravity darkening depends upon the underlying energy transport, the power-law scaling of gravity darkening (e.g., eq.~\eqref{eq:vonzeipel}) may be latitudinally dependent.  Prescriptions for modeling the latitudinal depth of convection are needed to investigate chemical mixing and magnetic field generation over the stellar surface of rapidly rotating stars \citep[e.g.,][]{kochukhov_bagnulo2006}.

\section{\label{sec:SSE} Solving the stellar structure equations}

\subsection{A brief overview of 1D stellar structure}

A wide variety of textbooks introduce the nominal stellar structure equations \citep[e.g.,][]{kippenhahn2013, hansen2004, lamerslevesque2017, Pinsonneault_Ryden2023} and some familiarity with these can maximize the value and interpretation of ESTER's predictions.  The standard equations of stellar structure consist of four first-order ordinary differential equations describing how the temperature $T_r$, the pressure $P_r$, the mass enclosed in a shell of radius $r$, $m_r$, and luminosity at a given radius, $L_r$, change from the center to the surface of the star\footnote{The subscript $r$ is added to remind the reader that these quantities are radius dependent, a redundancy that is beneficial, pedagogically.}.
In these equations, the radius $r$ (a spatial coordinate) is not explicitly dependent on time, making these an Eulerian form of the structure equations.  However, most graduate-level textbooks derive and describe these four equations using the mass, $m_r$, as the independent variable instead of $r$. This is useful since the mass of a star changes little over its (main sequence) lifetime while the radius can change substantially in time.  With $m_r$ as the independent variable the equations are Lagrangian and the radius explicitly becomes a dependent variable that can change in time, $r = r(t)$.  We point out these two approaches because ESTER v1.1.0rc2 uses the radius $r$ as the independent variable in its calculations.

In Eulerian form, the 1D stellar structure equations are
\begin{eqnarray} \label{hydro}
\frac{dP_r}{dr} &=&\frac{-Gm_r\rho}{r^2}~, 
\\ \label{masseq}
\frac{dm_r}{dr} &=& 4\pi r^{2}\rho~, 
\\ \label{eneq}
\frac{dL_r}{dr} &=&4\pi r^{2}\rho (\epsilon -\epsilon _{\nu })~,
\\ \label{entranseq}
\frac{dT_r}{dr} &=&-{3\kappa \rho L_r \over 64\pi r^{2}\sigma T_r^{3}} 
\hspace{0.1in} \mathsf{or} \hspace{0.1in} \frac{dT_r}{dr} =\left(1-{1 \over \gamma }\right){T_r \over P_r}{{\mbox{d}}P_r \over {\mbox{d}}r}~.
\end{eqnarray}
The first of these, eq.~\eqref{hydro}, is the equation of hydrostatic equilibrium that introduces the density $\rho$. The second equation, eq.~\eqref{masseq}, is the mass equation; spherical symmetry is assumed.  The third equation, eq.~\eqref{eneq}, is the energy equation, where the luminosity, $L_r$ (the energy flowing through a sphere of radius $r$), is determined by the energy generated from nuclear reactions, $\epsilon$, and the energy lost from
neutrinos, $\epsilon_{\nu}$.
The luminosity $L_r$ is zero at the center and rises to a constant value of $L$, the luminosity at the photosphere. The last equation, eq.~\eqref{entranseq}, is the equation of energy transport.  There are two versions of this equation, depending on whether energy is transported via radiation (the left version) or via convection (the right version).  The equation of radiative energy transport requires knowledge of the opacity, $\kappa$, while the equation of convective energy transport requires knowledge of the adiabatic index, $\gamma$, under the assumption of efficient convection.

To close these equations requires outside knowledge of three quantities: density $\rho$, nuclear energy generation rate $\epsilon$, and either the opacity $\kappa$ or adiabatic index $\gamma$.  In combination with the variables in the 4 stellar structure equations, we thus have seven dependent variables in total.  Three additional equations are needed to close the system. In stellar structure and evolution textbooks these additional equations as often called auxiliary equations. In practice, tables are used to look up the appropriate values of these constants for different states (i.e., pressure and temperature). The kind of opacity currently used for stellar structure calculations is the Rosseland mean opacity \citep{woods2024primer,farag2024expanded,seaton2004comparison}.  It is worthwhile to note here that a mode detailed multi-group opacity is already being used in the supernova community \citep[e.g.,][]{morag2025shock}.  Density is related to temperature and pressure via an equation of state (EOS), and this too is often interpolated from tabulated values that typically accompany the opacity calculations \citep{johnson1994sesame}. Textbooks usually approximate the EOS as an ideal gas because it is the simplest example. However, an ideal gas EOS is not relevant to stellar plasmas. To calculate $\epsilon$, analytic expressions are often used to estimate energy production from the P-P chain, the CNO cycle, and other reaction networks.  These estimates account for  energy lost via neutrinos, $\epsilon_{\nu}$ which is small for the P-P reactions (e.g., about 2\%), but is more substantial for CNO reactions \citep[e.g., as discussed by][]{kippenhahn2013}.

Finally, with four boundary conditions the four structure equations can be solved to determine how the seven dependent variables change with the independent variable radius.  The specifics of the boundary conditions will be discussed further in Section~\ref{sec:BCs}.  These solution variables can then be used to calculate other quantities of interest, such as the degree of ionization or the presence of convection at different radii.

\subsection{The ESTER equations for 1D and 2D stellar structure calculations}

While the four stellar structure equations described above underpin nearly all 1D stellar structure and evolution calculations, extending them to two and three dimensions is less natural.  There are a couple of reasons for this: (1) these 1D equations have the multi-dimensional fluid effects removed, and (2) the way they are written makes them difficult to solve numerically; equations of this type are referred to as ``stiff''~\footnote{Stiffness is a property of equations that makes them difficult to solve numerically \citep[e.g., as discussed by ][]{shampine1979user}.  We also refer to \url {https://en.wikipedia.org/wiki/Stiff_equation}.  Often one discretizes the equations and puts them in a matrix for numerical solution, and then one finds that that matrix is hard to invert, i.e. stiff.  This does not mean that the equations are unsolvable, but that more advanced numerical methods need to be applied.  A typical method is to employ a suitable preconditioner, which puts the matrix in a more pleasant form before one tries to invert it. } \citep{espinosa_lara_rieutord2013}.

The ESTER code computes solutions for stellar structures using a set of four equations, that we refer to as \emph{steady-flow stellar structure equations}.  This description indicates that a velocity, i.e. a steady flow, is included, but otherwise the equations are similar to hydrostatics in that they do not include time derivatives.   These four equations are described in \citet{espinosa_lara_rieutord2013, rieutord2016algorithm}:
\begin{eqnarray}\label{gravpoteq}
    \nabla^2 \phi &=& 4 \pi G \rho~,
    \\ \label{tempeq}
  \rho T \vec{v} \cdot \nabla S &=& - \nabla \cdot F_{\mathsf{heat,tot}} + \epsilon_{\mathsf{nuc}}~,
    \\ \label{momentumeq}
    \rho \vec{v} \cdot \nabla \vec{v} &=& - \nabla p - \rho \nabla \phi + F_v~,
    \\ \label{continuityeq}
    \nabla \cdot (\rho \vec{v}) &=& 0 ~.
\end{eqnarray}
Eq.~\eqref{gravpoteq} is the equation for the gravitational potential $\phi$, commonly known as Poisson's equation.  Eq.~\eqref{tempeq} describes the temperature field $T$, and includes the entropy, $S$. In this equation, $\epsilon_{\mathsf{nuc}}$ is the net sum of the energy generated and lost via neutrinos from nuclear reactions.  The total heat flux, $F_{\mathsf{heat,tot}}$, is calculated as 
\begin{eqnarray}
F_{\mathsf{heat,tot}} = -\chi \nabla T + F_{\mathsf{conv}}~.
\end{eqnarray}
Here $\chi$ is the thermal conductivity and $\nabla T$ is the temperature gradient.  The additional term, $F_{\mathsf{conv}}$,  is the energy flux due to convection, commonly known as the convective or enthalpy flux; the convective flux is obtained from a model for convection, and such a model is not included in this version of ESTER.
  Eq.~\eqref{momentumeq} derives from the momentum equation of hydrodynamics, in the {steady-flow} limit.  In this equation, $F_v$ is the viscous force and $p$ is the pressure. Eq.~\eqref{continuityeq} is the continuity equation of hydrodynamics in the {steady-flow case}.
  
Although the {steady-flow} stellar structure equations look very different from the 1D stellar structure equations, the goal is to determine how the same four stellar variables, $T_r$, $P_r$, $m_r$, and $L_r$ vary with radius.   This requires the same three auxiliary equations for the opacity, the nuclear energy generation, and the EOS.  These auxiliary equations and relevant physics are usually referred to as ``microphysics'' in the ESTER documentation \citep[e.g., Section 2.3 in][]{rieutord2016algorithm}; this is a common term in the hydrodynamics community,  For example, in the climate community, microphysics could also include moisture models and other inputs that affect the smallest scale motions  \citep[e.g.,][]{Keller2023,van2013role}, and that are different from the microphysics relevant for stars.

\subsection{\label{sec:BCs} Boundary conditions}

\begin{figure}[h]
\begin{center}
\resizebox{6in}{!}{\includegraphics{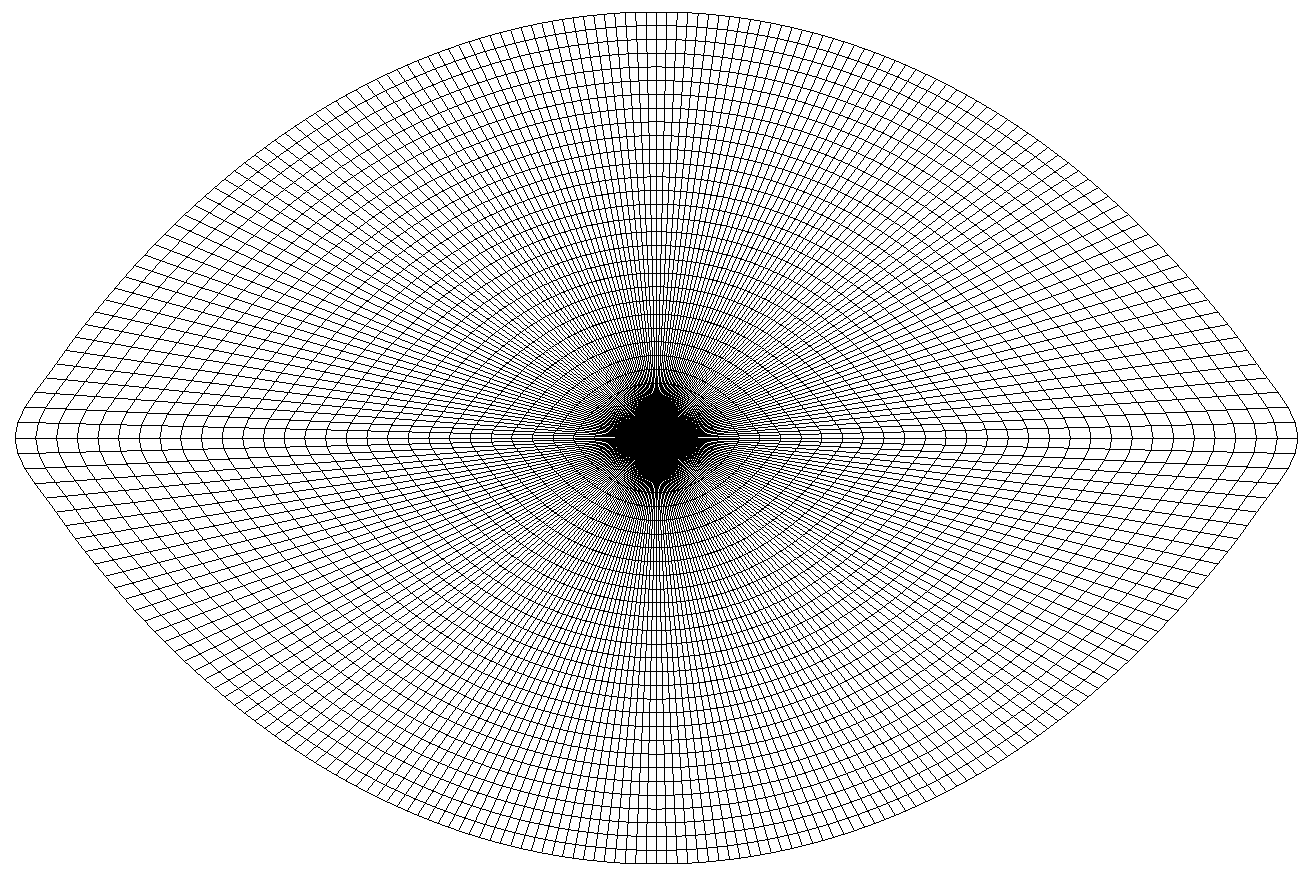}}
\caption{The computational grid used in a 2D ESTER stellar structure model illustrating the spatial discretization used in the calculation.
\label{fig:estergrid} }
\end{center}
\end{figure}

The four stellar structure equations (whether the standard formulation or the steady-flow formulation in ESTER), require four boundary conditions so that they can be solved over the domain of the stellar interior. These boundary conditions can be imposed at the center and/or at the surface of the star.  For the standard form of the stellar structure equations that evolve quantities between the center of the star and the surface, imposing conditions at the center of the star is natural.  Boundary conditions on mass and luminosity are typically imposed so that $m_r(r=0) = 0$ and $L_r(r=0) = 0$.   Central boundary conditions are also used in ESTER calculations.  Figure \ref{fig:estergrid} illustrates an ESTER computational grid for the interior of a rapidly rotating star.  The grid consists of elements with four corners (i.e., irregular quadrilaterals), but these shapes cannot tessellate a sphere without a degenerate point.  That degeneracy happens at the center of the star, where many elements that would otherwise have four corners effectively have 3 corners.  Numerical instabilities arise from this kind of mesh.  The ESTER documentation states that they impose conditions such that ``the solutions should simply be regular'' \citep{espinosa_lara_rieutord2013} and ``we just need to impose the regularity of the fields'' \citep{rieutord2016algorithm}.  The solutions are required to be smooth and well-behaved at the degenerate center point.

The remaining boundary conditions are imposed at the surface of the star. The first is that the gravitational potential vanishes at infinity.  \citet{espinosa_lara_rieutord2013} provide a qualitative description of the ``self-consistent field method'' that is adopted to achieve this.   Two additional boundary conditions are used to achieve stress-free conditions on the velocity field.  These are
\begin{eqnarray} \label{reflectbceq}
\vec{v} \cdot \vec{n} = 0~,
\\ \label{streefreebceq}
([\sigma]\vec{n}) \wedge \vec{n} = \vec{0}~,
\end{eqnarray}
where [$\sigma$] is the stress tensor and $\vec{n}$ is the outer normal of the star.  Qualitatively, the fluid is to feel no ``horizontal'' stress, i.e. eq.~\eqref{streefreebceq} requires that the stress in the direction perpendicular to the boundary be zero.  The fluid must also remain confined to the star; this is achieved in eq.~\eqref{reflectbceq}, which provides a boundary condition that is referred to as ``reflective''; at the boundary, the velocity in the direction of the surface normal is zero.  Finally, a boundary condition is imposed on the temperature field.  It is assumed that the star radiates locally as a black body
\begin{eqnarray}
-\chi_r\vec{n} \cdot \nabla T = \sigma_{\mathsf{SB}} T^4.
\end{eqnarray}
Colloquially this is often referred to as ``sigma-T-4'' radiation.
The velocity and temperature boundary conditions are applied at the bounding surface of the star, defined in the following section.

\subsection{\label{sec:surface} The bounding surface and visible surface of the star in ESTER}

ESTER defines the bounding surface of the star to be a surface of constant pressure, calibrated at the pole and defined as
\begin{eqnarray} \label{Presssure_BC}
P_{\mathsf{s}} = \tau_{\mathsf{s}} {g_{\mathsf{pole}} \over \kappa_{\mathsf{pole}}}~.
\end{eqnarray}
Here the subscript ``s'' indicates bounding surface values and $\tau_s$ indicates the optical depth.  As noted above in Section \ref{sec:BCs}, this bounding surface is where the boundary conditions are imposed.  All stellar parameter profiles calculated by ESTER run from the central value to this bounding surface.

The bounding surface of the star is \textit{not} the photosphere, the visible surface of the star at which the temperature is the effective temperature, $T_{\mathsf{eff}}$.  The photosphere occurs where $\tau_{\mathsf{s}} = 2/3$, or approximated as $\tau_{\mathsf{s}} = 1.0$ in the ESTER code.  In a rotationally distorted star, this optical depth will occur at different pressures at different latitudes, and thus will not occur at the (constant pressure) bounding surface, nor at a constant distance from the bounding surface.  The exception is the case where the star is not rotating and then temperature and $T_{\mathsf{eff}}$ are equal everywhere on the bounding surface.  The ESTER code calculates the temperature at $\tau_{\mathsf{s}} = 1.0$ (the photosphere) by extrapolating the temperature from the bounding surface to this visible
surface.

\subsection{The effective temperature surface of the star in ESTER \label{sec:teff}}

An important success of the ESTER code has been its ability to predict the light distribution across the visible surface of rapidly rotating stars that can be compared with observations (e.g., Figure \ref{fig:rapid_rotators}). To accomplish this in practice requires extending the temperature profiles (that are truncated at the bounding surface) to the visible surface (the photosphere) where the temperature is $T_{\mathsf{eff}}$. The bounding surface has a temperature equal to $T_{\mathsf{eff}}$ only at the pole. At other latitudes the bounding surface is below the visible surface, and $T_{\mathsf{b}}(\theta)$ is larger than $T_{\mathsf{eff}}$.  

ESTER accomplishes this extrapolation in temperature to the visible surface by modeling the star's atmosphere as a polytrope of index 3 ($n=3$).  Polytropes provide an analytical way to model a self-gravitating gaseous sphere \citep{chandrasekhar1967}. A polytropic relation between pressure and density is typically expressed as
\begin{eqnarray} \label{eq:polytrope}
P = K \rho^{1 + {1 \over n}}~.
\end{eqnarray}
Because it relates the pressure and density, this equation is also an example of an analytic equation of state (EOS).  \citet{espinosa_lara_rieutord2013} discuss the polytropic relation of temperature as a model atmosphere, and point out that ``a more realistic atmospheric model can be used".  For example, relations between density and pressure could be obtained from tabulated values.  This could be useful for investigating the conditions in atmospheres more completely (e.g., convective stability; ionization fraction).  Their choice of an $n=3$ is a common representation for radiative regions of main sequence stars
\citep[e.g., Eddington's Standard Model;][]{eddington1926}.

For the present discussion, \citet{espinosa_lara_rieutord2013}'s equation for the effective temperature at any latitude, represented by the angle $\theta$,
\begin{eqnarray} \label{eq:teff}
T_{\mathsf{eff}} = T_{\mathsf{b}}(\theta) \left[ {g_{\mathsf{eff}}(\theta) \over g_{\mathsf{pole}}} {\kappa_{\mathsf{pole}} \over \kappa(\theta)} \right]~,
\end{eqnarray}
needs to be highlighted.  The \emph{visible surface} temperature, the true $T_{\mathsf{eff}}$, can be calculated from values on the bounding surface ($T_{\mathsf{b}}(\theta)$, $g_{\mathsf{eff}}(\theta)$,$\kappa(\theta)$) and polar values ($g_{\mathsf{pole}}$,
$\kappa_{\mathsf{pole}}$).  Since the light distribution across a star is set by the temperature distribution across the star, this formalism enables an ``image'' of a rotationally distorted star to be created.

\subsection{Setting the angular velocity}

ESTER requires as an input value the angular velocity $\omega$, as a fraction of the critical angular velocity of the star $\omega_{\mathsf{crit}}$.  By specifying the angular velocity of the star, the equations are fully constrained.  While this can also be done by specifying the total angular momentum of the star, 
\begin{eqnarray} \label{eq:angular_momentum}
L_{\mathsf{ang}} = \int r \text{sin} \theta \rho u_\phi {\mbox{d}}V,
\end{eqnarray}
in practice it is more common to estimate the equatorial velocity, $V_{\mathsf{eq}}$, from observations.  The total angular momentum of the star is then constrained by specifying
\begin{eqnarray} \label{eq:equatorial_vel}
V_{\mathsf{eq}} = v_\phi(r = R, \theta = \pi/2).
\end{eqnarray}
A projection of the equatorial rotational velocity of a star, called the projected rotational velocity or $V_{\mathsf{eq}} \sin{i}$, can be measured from the width of a star's spectral lines in high dispersion spectra.  Some translation is therefore required from observational calculations to inputs for ESTER and other stellar structure codes, using. e.g., eq.~\eqref{transformtoangularmomentum}.

\section{\label{sec:Profiles} Stellar variables calculated via ESTER}

An understanding of the interior properties and governing physics of a star can be gained by investigating how stellar structure variables change through the star's interior. We refer to plots of a structure variable versus radius, or versus some tracer of radius (e.g., mass interior to a radius, temperature, etc.) as \emph{radial profiles} or simply \emph{profiles}. For rotating stars that are oblate in shape, the stellar structure variables are a function of both the radial coordinate, $r$, and the angular coordinate that spans between the pole and the equator, $\theta$. Notationally, $\theta = 0 \degree$ at the pole and $\theta = 90\degree$ at the equator. In the default values of the input parameters that we use here, the ESTER code calculates radial profiles with a sampling of 900 units, running from the center to the bounding surface of the star.   We investigate and illustrate several stellar structure variables calculated by the ESTER code for stars that have been distorted by rapid rotation.  This exploration is not exhaustive, and reflects our own interests. The star's rate of rotation is quantified relative to its critical angular velocity. Some redundancy of terms and definitions is included in these descriptions; this guide is structured to be a reference rather than a continuous monograph.

\subsection{Density, $\rho$, and Pressure, $P$}
\begin{figure}[h]
\begin{minipage}[c]{0.65\linewidth}
\includegraphics[width=\linewidth]{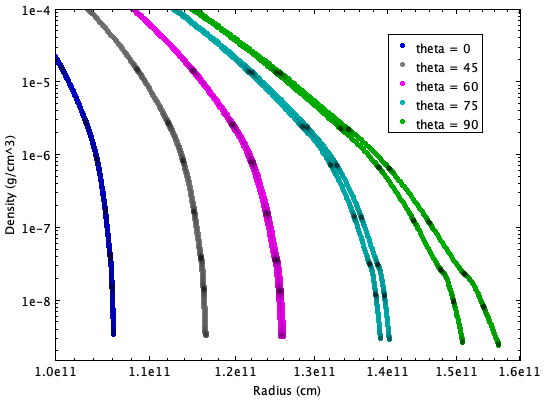}
\end{minipage}
\begin{minipage}[c]{0.35\linewidth}
 \caption{
 Density versus radius for two stars, both with a mass of $1.86\,M_\odot$, rotating at 95\% and 98\% of the critical rotation rate. Density profiles for each are shown at $\theta$ values of $0^{\degree}$ (pole), $45^{\degree}$, $60^{\degree}$, $75^{\degree}$ and $90^{\degree}$ (equator). Increased rotation causes the radius of the star to become larger toward the equator and the surface to be less dense at the equator. \label{fig:density_radius} 
 }
\end{minipage}
\end{figure}
An equation of state provides the relationship between the density, pressure, and temperature.  Figure \ref{fig:density_radius} illustrates density versus radius for two stellar structures produced with ESTER, both with a mass of $1.86\,M_\odot$, rotating at 95\% and 98\% of their critical rotation rate respectively.  Density profiles for each are shown over a range of $\theta$ values, spanning from the pole ($0 \degree$) to the equator ($90 \degree$).  Each profile runs from the center
to the bounding surface of the star, however the figure displays a radial range near the bounding surface so that the differences in these densities are visible.  Faster rotation results in 
the star's radius being larger as $\theta$ is increased from the pole to the equator.  Increased rotation also causes the density to drop as $\theta$ is increased from the pole to the equator.

\begin{figure}[h]
\begin{minipage}[c]{0.5\linewidth}
\includegraphics[width=\linewidth]{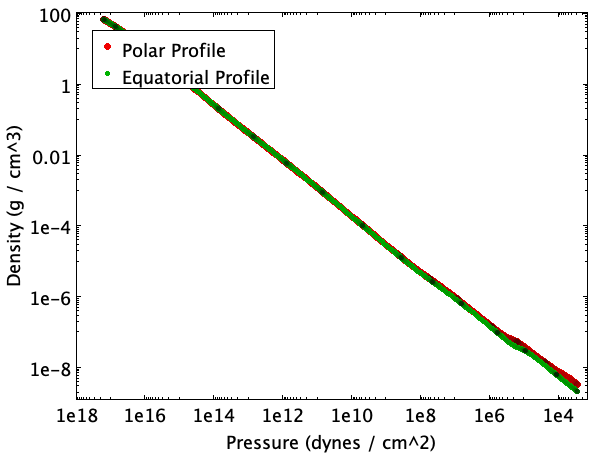}
\end{minipage}
\begin{minipage}[c]{0.5\linewidth}
\includegraphics[width=\linewidth]{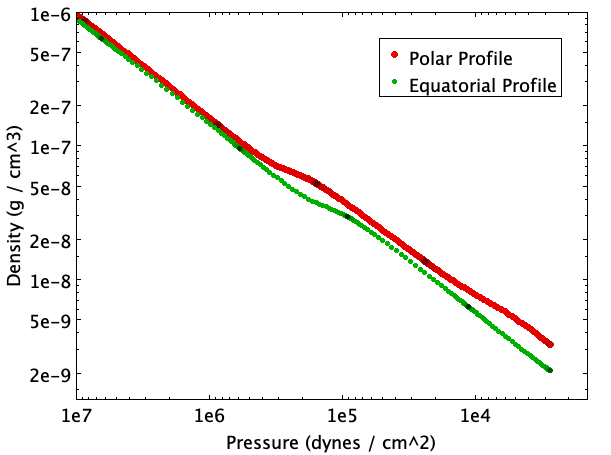}
\end{minipage}%
      \caption{Density versus pressure in cgs units for a $1.4\,M_\odot$ star rotating at 98\% of the critical angular velocity; both polar and equatorial profiles are plotted.  The \textit{right panel} shows the full range of density while the \textit{left panel} shows the outer $\sim 3$ orders of magnitude.  Density values are identical at the center but diverge toward the surface. The distorted shape causes the density to decrease near
      the equator (at a given pressure). Note that the polar and equatorial profiles extend to the same pressure (at the bounding surface), but different physical sizes, because the star is rotationally distorted.
      \label{fig:density_pressure} 
      }
\end{figure}

To illustrate the relationship between density and pressure, Figure \ref{fig:density_pressure} shows polar and equatorial profiles of density versus pressure for a $1.4\,M_\odot$ star rotating at 98\% critical angular velocity.  The values of pressure and density for all values of $\theta$ are identical at the center.  Following these profiles from the center toward the bounding surface, the density of equatorial profile becomes less than the polar profile, a consequence of rotational distortion.
Since the bounding surface of an ESTER model is a surface of constant pressure (see Section \ref{sec:surface}), the last value of pressure (at the bounding surface) will be the same at all values of $\theta$, as the plot shows, even though they correspond to different radii if the star is oblate because of rotation. Since the density is lower at the equator, pressure equilibrium is maintained by increasing the temperature. The bounding surface is \emph{hotter} at the equator than at the poles. This is contrary to what is expected for gravity darkening; recall that this bounding surface is not the visible surface of the star at the photosphere.

\subsection{Temperature, $T$}

\begin{figure}[h]
\resizebox{6in}{!}{\includegraphics{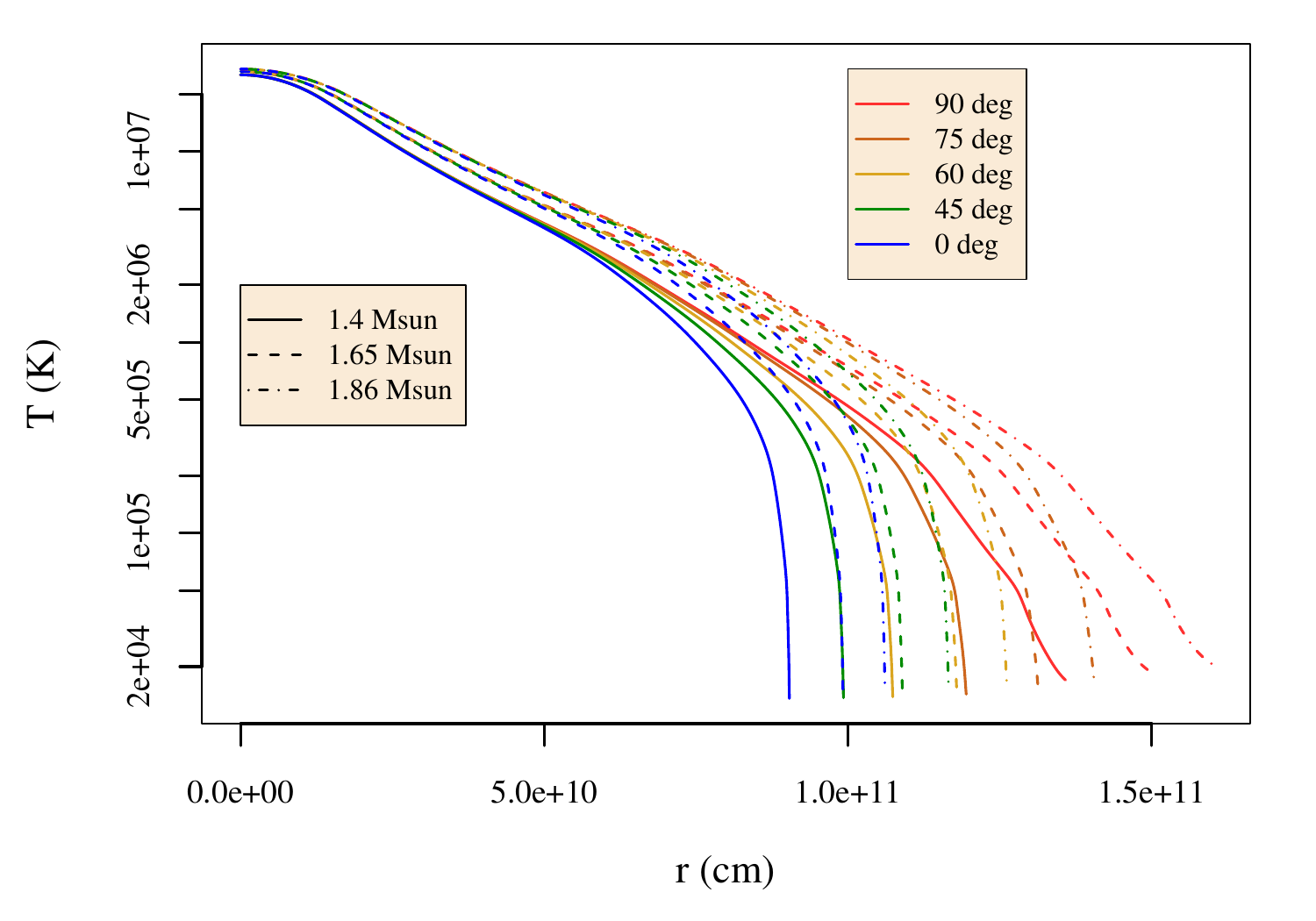}}
      \caption{Temperature vs radius for three stars rotating at 98\% of the critical angular velocity
      with masses of $1.4\,M_\odot$, $1.65\,M_\odot$ and $1.86\,M_\odot$.   A range of
      profiles are shown
      between the polar axis ($\theta = 0 \degree$) and the equator ($\theta = 90 \degree$).
      These rapidly rotating stars have larger equatorial radii than polar radii, and
      the equatorial radii become larger with increased rotation; the polar profiles
      do not change with rotation, and are overlapping.  However, the equatorial 
      temperature value at the bounding surface \textit{increases} with rotation.
      \label{fig:temp_versus_radius}}
\end{figure}

Figure \ref{fig:temp_versus_radius} illustrates the temperature profile of three rapidly rotating stars with masses of $1.4\,M_\odot$, $1.65\,M_\odot$ and $1.86\,M_\odot$.  All three are rotating at 98\% of their critical angular velocity.  For each star, profiles at five different latitudes ($\theta = 0^{\degree}$, $45^{\degree}$, $60^{\degree}$, $75^{\degree}$ and $90^{\degree}$) are plotted.  The temperature profiles for these stars are nearly independent of $\theta$ from the center through roughly the central one-third of the star. At distances further from the center, the temperature profiles diverge, with profiles closer to the equator remaining hotter than those closer to the pole. The profiles terminate at the bounding surface that is both larger and hotter at the equator than at the pole. Here it is also important to recall that the temperature of the bounding surface of the ESTER model is not the temperature of the star's visible surface.

\subsection{Surface Energy Flux, $F_{\mathsf{surf}}$, and Effective Temperature, $T_{\mathsf{eff}}$ \label{sec:teffviz}}

The effective temperature of a star is defined to be the temperature of a blackbody that emits the same power per unit area. A spherical star is uniformly bright across its surface and therefore has the same effective temperature across its surface. This picture neglects small-scale features like star spots, cooler local areas that are caused by a specific magnetic field configuration that frequently occurs on the surface of young, active stars \citep[e.g.,][]{paolino2025,borrero2011magnetic,solanki2003sunspots}. For rapidly rotating stars, the effect of gravity darkening results in stars having a higher surface energy flux, $F_{\mathsf{surf}}$, and higher effective temperatures at the poles than at their equators.  

The product of the \textit{average} surface energy flux, $ \langle F_{\mathsf{surf}} \rangle$, 
and the total surface area, $A$, always equals the star's luminosity, $ A \langle F_{\mathsf{surf}}\rangle
= L$.  For a spherical star, $A = 4 \pi R^2$; for an oblate spheroid, the area can be calculated
from the semi-major and semi-minor axes. Rapidly rotating stars are not strictly oblate spheroids, but regardless of the geometry, this relation can be combined with the Stefan-Boltzmann law to define {an average} effective temperature using,
\begin{eqnarray}
L =  A \sigma_{\mathsf{SB}} T_{\mathsf{eff}}^4 ~,
\end{eqnarray}
where $\sigma_{\mathsf{SB}}$ is the Stefan-Boltzmann constant.
Thus we see that calculating the area is not necessary to define the local
effective temperature as, $T_{\mathsf{eff}}^4 = \langle F_{\mathsf{surf}} \rangle / \sigma_{\mathsf{SB}}$.

\begin{figure}
\begin{center}
\resizebox{5in}{!}{\includegraphics{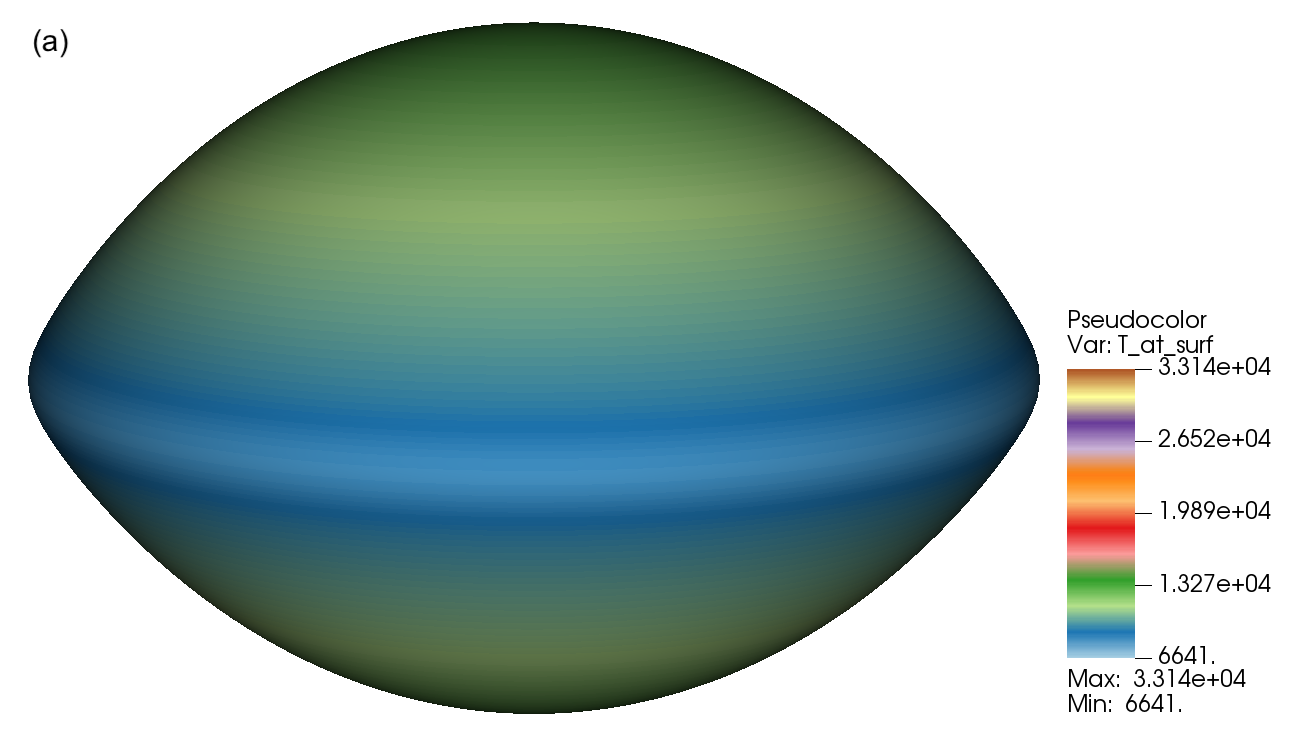}}
\resizebox{5in}{!}{\includegraphics{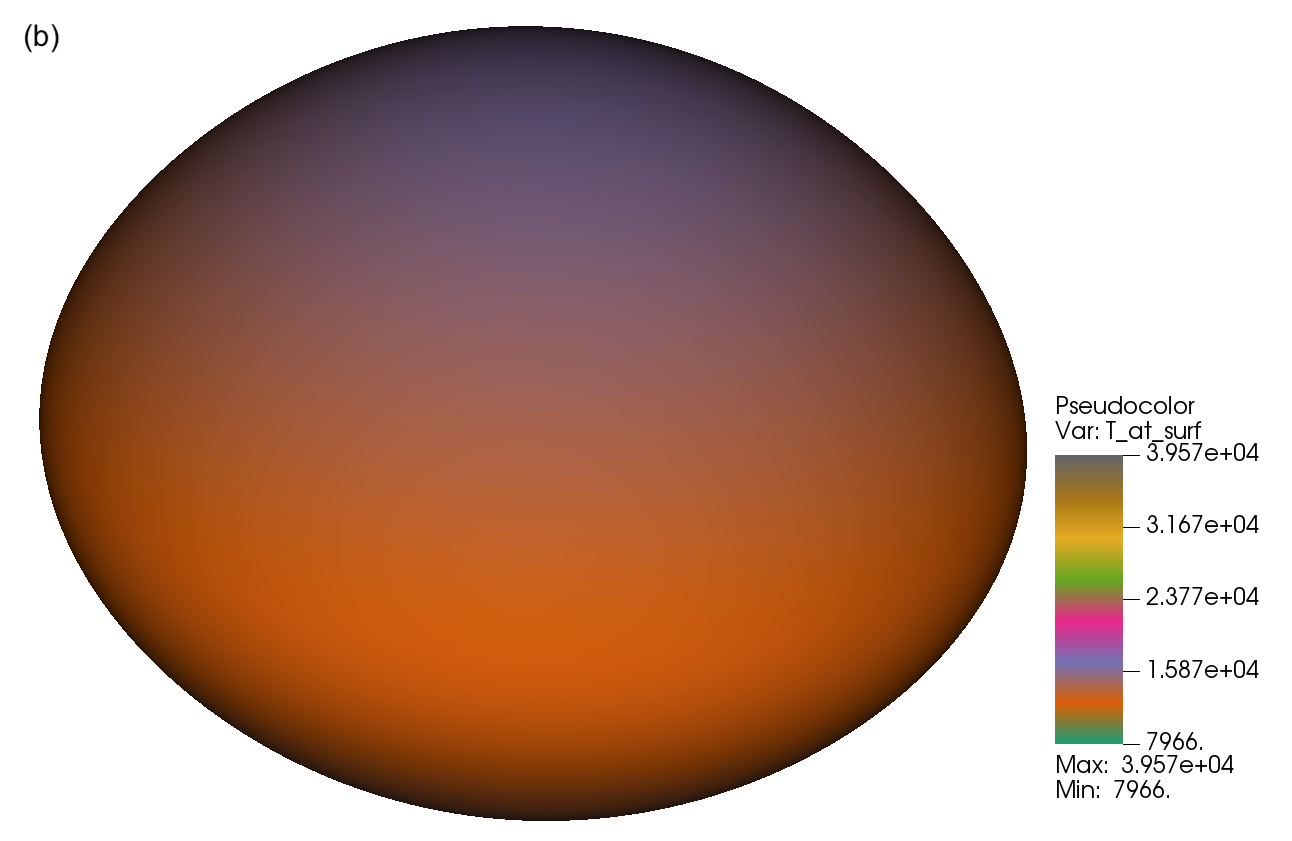}}
\caption{Visualization of the effective temperature on the visible surface of a 1.4 $M_{\odot}$ star (a) rotating at 95\% of the critical angular velocity, and a 1.86 $M_{\odot}$ star (b) rotating at 80\% of the critical angular velocity. The two visualizations are not on the same size scale. These visualizations illustrate that the phenomenon of gravity darkening, caused by cooler temperatures at the equator, is reproduced in ESTER stellar structure models.
\label{fig:esterviztemp} }
\end{center}
\end{figure}

As described in Section \ref{sec:teff}, the ESTER code calculates effective temperature
at the visible surface of the star, defined to be where $\tau = 1$, using a
polytropic extension above the bounding surface. Figure~\ref{fig:esterviztemp} illustrates
the effective temperature over the surface of 2 rapidly rotating stars produced using ESTER. These stars 
clearly exhibit the expected gravity darkening: their equators are cooler than their poles. These
visualizations should not be directly compared to images of rapidly rotating stars (e.g.,
Figure \ref{fig:rapid_rotators}); instead, comparisons should be made to the local surface
flux, $F_{\mathsf{surf}}(\theta) = \sigma_{\mathsf{SB}} T_{\mathsf{eff}}(\theta)^4$, at the wavelength of
the observations.
To quantify the extent of the temperature extrapolation from the bounding surface to the 
visible surface, in Table~\ref{tab:2msun_temperatures} we list bounding surface temperatures
and effective temperatures for a $2\,M_\odot$ star at five different values of $\theta$ and
rotating at three rates of rotation; the radial extent of the bounding surface is also provided.
For these stars, relative to the non-rotating model, rotation causes the polar radius
to shrink by 1.0\% at $\omega/\omega_{\mathsf{crit}}$ = 0.60 and by 1.4\% at $\omega/\omega_{\mathsf{crit}}$ = 0.95.  And
correspondingly, the polar effective temperature increases by 3.3\% at $\omega/\omega_{\mathsf{crit}}$ = 0.60
and by 4.6\% at $\omega/\omega_{\mathsf{crit}}$ = 0.95. More dramatic changes occur at the equator, as expected.
Rotation causes the equatorial radius to increase by 17.0\% at $\omega/\omega_{\mathsf{crit}}$ = 0.60 and 
by 46.7\% at $\omega/\omega_{\mathsf{crit}}$ = 0.95.  And the equatorial effective temperature decreases by 
11.9\% at $\omega/\omega_{\mathsf{crit}}$ = 0.60 and by 33.1\% at $\omega/\omega_{\mathsf{crit}}$ = 0.95.

\begin{table}
\rowcolors{2}{gray!10}{white}
\centering
\caption{Sizes and temperatures of a rotating $2\,M_\odot$ star.  The radius at the bounding surface $R_{\mathsf{b}}$, the temperature at the bounding surface $T_{\mathsf{b}}$, and the
effective temperature on the visible surface $T_{\mathsf{eff}}$ are provided.
  These values can be compared to those of a non-rotating $2\,M_\odot$ star (also calculated by ESTER in its
1D mode) that has $T_{\mathsf{b}} = 8896$ K, effective temperature $T_{\mathsf{eff}} = 8896$ K, and $R_{\mathsf{b}} = 1.277 \times 10^{11}$ cm.}
\begin{tabular}{cccccccccc}
& \multicolumn{3}{c}{$\omega/\omega_{\mathsf{cr}}$ = 0.60} &
\multicolumn{3}{c}{$\omega/\omega_{\mathsf{cr}}$ = 0.80} & 
\multicolumn{3}{c}{$\omega/\omega_{\mathsf{cr}}$ = 0.95}
\\
$\theta$ &
$R_{\mathsf{b}}$ & $T_{\mathsf{b}}$ & $T_{\mathsf{eff}}$ &
$R_{\mathsf{b}}$ & $T_{\mathsf{b}}$ & $T_{\mathsf{eff}}$ & 
$R_{\mathsf{b}}$ & $T_{\mathsf{b}}$ & $T_{\mathsf{eff}}$
\\
(deg) &
($10^{11}$ cm) & (K) & (K) &
($10^{11}$ cm) & (K) & (K) &
($10^{11}$ cm) & (K) & (K)
\\
\hline
0   & 1.264 & 9192 & 9192    & 1.260 & 9273 & 9273      &  1.259 & 9301 & 9306 \\
45  & 1.348 & 9124 & 8719    & 1.374 & 9147 & 8684      &  1.383 & 9152 & 8678 \\
60  & 1.408 & 9081 & 8359    & 1.471 & 9032 & 8176      &  1.496 & 8996 & 8127 \\
75  & 1.467 & 9032 & 8002    & 1.593 & 8840 & 7522      &  1.664 & 8658 & 7330 \\
90  & 1.494 & 9006 & 7835    & 1.671 & 8685 & 7018      &  1.873 & 7919 & 5949 \\
\hline
\end{tabular}
\label{tab:2msun_temperatures}
\end{table}

\subsection{Opacity, $\kappa$ \label{sec:opac} } 

Opacity, or the mass absorption coefficient, is a measure of the impenetrability to electromagnetic radiation.  For main sequence stars, the opacity is generally lowest in the fully ionized interior of a star, being dominated by electron scattering and thus nearly independent of density and temperature. Toward the cooler outer layers of a star ($T < 10^6$ K), the opacity increases dramatically as metals, helium and then hydrogen transition to being partly ionized. This increase in opacity causes convection in low mass stars and red giant stars. Past this partial ionization zone, the opacity drops toward zero at the visible surface. ESTER calculates opacities by interpolating tables generated via the Opacity Project at Livermore (OPAL) \citep{rogers1996opal,rogers1994opal,iglesias1996updated}. We note that updated and improved tables now exist, including the OPLIB tables \citep{farag2024expanded} and the Livermore opacity libraries \citep{grabowski2021opacity,aberg2020opus}. These could be used in a future version of the code along with the SESAME equation of state database \citep{mchardy2018introduction} or LEOS library \citep{macfarland1999new}.

\begin{figure}[h]
\begin{minipage}[c]{0.5\linewidth}
\includegraphics[width=\linewidth]{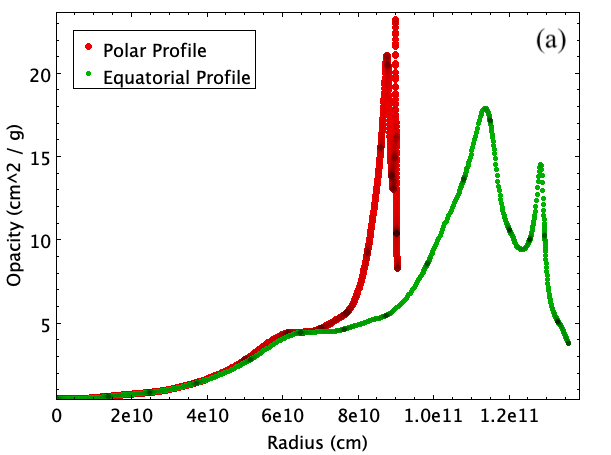}
\end{minipage}
\begin{minipage}[c]{0.5\linewidth}
\includegraphics[width=\linewidth]{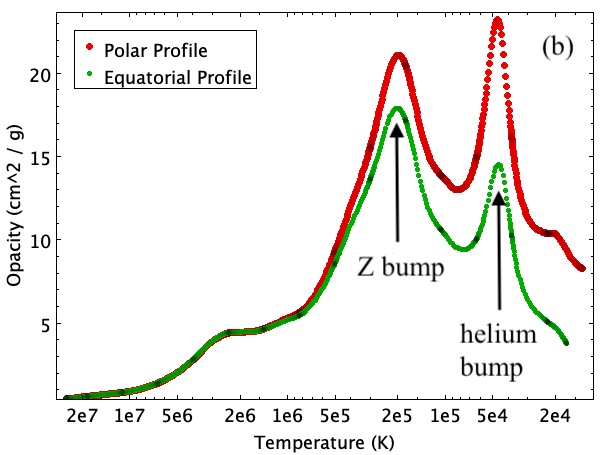}
\end{minipage}%
\caption{(\textit{a})  Opacity versus radius and (\textit{b}) opacity versus temperature 
 for a $1.4\,M_\odot$ star rotating at 98\% of the critical angular velocity. The star has a larger equatorial radius than polar radius because of rapid rotation.  The peaks in opacity at cooler temperatures and nearing the surface are a consequence of the metal bump at $\sim 2 \cdot 10^5$ K and the second ionization of helium at $\sim 5 \cdot 10^4$ K \citep[e.g.,][]{guzik2018opacity}
\label{fig:opacity}
}
\end{figure}

Two profiles of opacity generated with the OPAL tables (versus radius and versus temperature) through a $1.4\,M_\odot$ star rotating at 98\% of the critical angular velocity are shown in Figure \ref{fig:opacity}; this includes profiles taken in the direction of the pole and the equator. Given the smaller physical extent of the polar profile, it encounters the partial ionization zone and larger opacities at a smaller radius than the equatorial profile does. The polar and equatorial opacity profiles have more similar shapes when they are plotted versus temperature; both show two peaks at cooler temperatures corresponding to the ionization temperature for helium and of hydrogen. The higher opacity of the polar profile at a given temperature stems from the slightly higher density profile (see Figure \ref{fig:density_pressure}).

\subsection{Local Effective Gravity, $g_{\mathsf{eff}}$}

The local effective gravity, $\vec{g}_{\mathsf{eff}}$, is the gravitational acceleration combined with the centrifugal acceleration, namely
\begin{eqnarray}\label{eqgeff}
    \vec{g}_{\mathsf{eff}} = -\nabla\phi+\omega^2(r,\theta)r\sin\theta\vec{e}_s~.
\end{eqnarray}
Here $\phi$ is the gravitational potential calculated using Poisson's equation discussed in Section~\ref{secgravpot}, $\omega(r,\theta)$ is the local angular velocity, and $\vec{e}_s$ the radial cylindrical unit vector.

\begin{figure}[h]
\begin{minipage}[c]{0.65\linewidth}
\includegraphics[width=\linewidth]{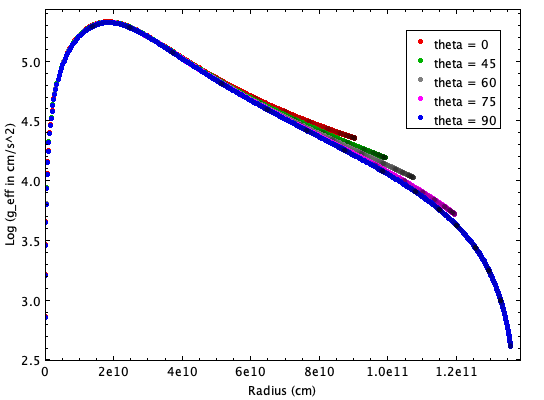}
\end{minipage}
\begin{minipage}[c]{0.35\linewidth}
      \caption{Local effective gravity in cgs units from the center to the bounding surface for a
      $1.4\,M_\odot$ star rotating at 98\% of the critical angular velocity. Five profiles are plotted ranging from 
      polar ($\theta = 0 \degree$) to equatorial ($\theta = 90 \degree$). Rotation causes the local
      gravity to decrease toward the equator, decreasing the effective temperature and leading to 
      the phenomenon of gravity darkening.
      \label{fig:geff_radius} }
\end{minipage}%
\end{figure}

The variation of $g_{\mathsf{eff}}$ with radius for a $1.4\,M_\odot$ star rotating at 98\% of the critical angular velocity is illustrated in Figure~\ref{fig:geff_radius}, over a range of $\theta$ values. A large drop of $g_{\mathsf{eff}}$ values toward the equator is evident. This demonstrates the strength of the centrifugal acceleration at the equator.  Because effective temperature roughly scales with the local effective gravity \citep[e.g.,][; eq.~ \eqref{eq:vonzeipel}]{vonzeipel1924} this results in a cooler, less luminous equatorial region and the phenomenon of gravity darkening.

\subsection{The Gravitational Potential, $\phi$ \label{secgravpot}}

\begin{figure}[h]
\begin{minipage}[c]{0.5\linewidth}
\includegraphics[width=\linewidth]{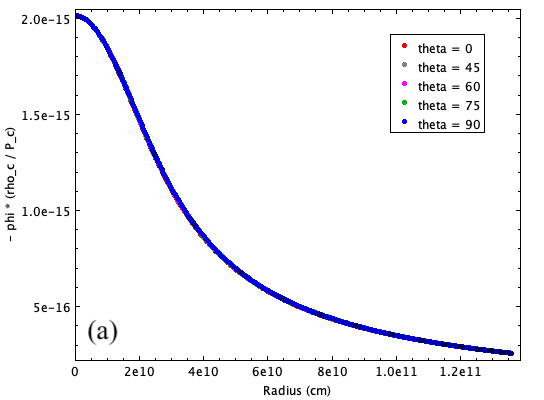}
\end{minipage}
\begin{minipage}[c]{0.5\linewidth}
\includegraphics[width=\linewidth]{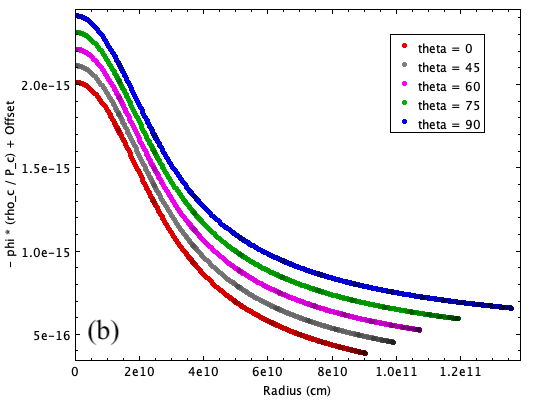}
\end{minipage}%
      \caption{The negative of the gravitational potential versus radius of a $1.4\,M_\odot$ star rotating at 98\% of the critical angular velocity. Profiles at five values of $\theta$, ranging from the pole ($\theta = 0^{\degree}$) to the equator ($\theta = 90^{\degree}$) are shown. ESTER computes a scaled version of the gravitational potential, $\phi \times P_{\mathsf{c}} / \rho_{\mathsf{c}}$,
      and that scaling is removed here. The full extent of these profiles is shown in (a) while (b) focuses on the termination of these surface profiles at the bounding surface; small offsets of $n \times 10^{-14}$ are added to separate the curves. The polar profile is more negative than those toward the equator because the mass interior is less spatially extended at the pole than near the equator.
      \label{fig:grav_pot_radius}}
\end{figure}

The Lagrangian form of the stellar structure equations uses $m_r$, the mass enclosed in radius $r$, as the independent variable and calculates other variables as a function of it (e.g., Section 2.1). But because rapidly rotating stars are {not} spherically {symmetric}, ESTER instead calculates the gravitational potential, $\phi$, through the star. This is used to solve Poisson's equation,
\begin{eqnarray}\label{gravpoteq2}
    \nabla^2 \phi &=& 4 \pi G \rho~,
\end{eqnarray}
 one of the four {steady-flow} stellar structure equations.
The gravitational potential value {calculated} by ESTER is scaled by the ratio of the central pressure to the central density, $P_{\mathsf{c}} / \rho_{\mathsf{c}}$ \citep[see Section \ref{sec:surface} in][]{rieutord2016}.  We remove this scaling for illustrative purposes and plot the (negative of the) gravitational potential versus radius for a $1.4\,M_\odot$ star rotating at 98\% of the critical angular velocity in Figure \ref{fig:grav_pot_radius}. The value is negative by convention. Near the surface, the polar profile of $\phi$ is more negative than those near the equator because the mass interior is less spatially extended at the pole than near the equator.

\section{\label{sec:cb} Boundaries of Convective Instability}

 The currently available ESTER code does not include a prescription for convection in the outer envelope of a star (see the discussion in Section \ref{sec:capab}). This is why ESTER is most suitable for modeling stars with radiative {envelopes}, typically main sequence stars more massive than $\sim 1.4\,M_\odot$. However, 1D stellar structure models predict that intermediate and higher mass stars can have small regions in their exterior that are formally convectively unstable \citep[e.g., Figure 2 in][]{cantiello_braithwaite2019}. Because these regions are thin, in practice, they have typically been ignored. In this section we explore and illustrate the layers of rapidly rotating stars that are formally unstable to convection as defined by 3 different criteria available in ESTER. Instability to convection is often described in stellar structure and evolution using the simple picture of a displaced parcel of stellar plasma.  If this parcel of fluid will rise (due to buoyancy) or sink, it will lead to the onset of convective motion.  In fluid dynamics, this is described by the Rayleigh number, the product of the Grashof and Prandtl numbers.  The Rayleigh number expresses the balance of the forces due to buoyancy and viscosity, as well as the balance between the kinematic diffusivity and thermal diffusivity.  For a given material (with thermal diffusivity, viscosity, and buoyancy properties), a critical Rayleigh number can be calculated.  Rayleigh numbers above this critical value, indicate that the temperature gradient will lead to convection.  Stellar astrophysics expresses this critical threshold for convection through the Schwarzschild criterion or the Ledoux criterion.

\subsection{The Adiabatic Gradient}

The Schwarzschild criterion for convection uses the temperature and pressure gradients along with the specific heats of a fluid to define convective instability.  Notationally, $\nabla =  d \ln T / d \ln P$ and $\nabla_{\mathsf{ad}} \sim 1 - 1 / \gamma$, with $\gamma$ equal to the adiabatic index, a heat capacity ratio ($c_P / c_V$). For example, a monatomic ideal gas has $\nabla_{\mathsf{ad}} = 0.4$. This is not strictly correct for stellar interiors, but is a common approximation in textbooks. To express the Schwarzschild criterion, we define the so-called Schwarzschild discriminant
\begin{eqnarray}
S(r) = \nabla_{\mathsf{ad}} - \nabla~.
\end{eqnarray}
The Schwarzschild criterion states that the
Schwarzschild discriminant, $S(r)$, must be greater than zero for stability against convection.  When $S(r)<0$, the convective instability will generate large-scale fluid motions. This Schwarzschild criterion for convection is sometimes referred to as superadiabaticity (i.e., the gradient $\nabla$ is greater than expected for an adiabatic displacement $\nabla_{\mathsf{ad}}$).  

\begin{figure}[h]
\begin{minipage}[c]{0.65\linewidth}
\includegraphics[width=\linewidth]{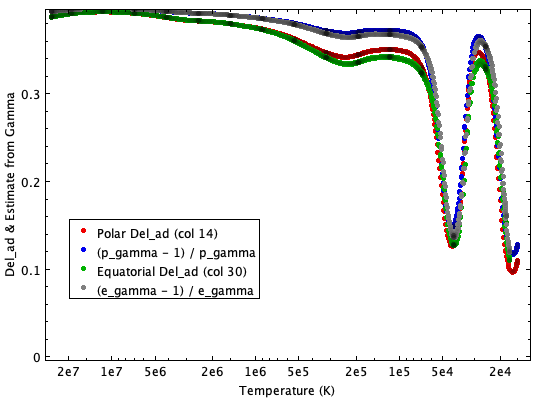}
\end{minipage}
\begin{minipage}[c]{0.35\linewidth}
      \caption{The adiabatic gradient $\nabla_{\mathsf{ad}}$ and an estimate of it, $(\gamma-1)/\gamma$, versus temperature 
      along the polar and equatorial axes for a $1.65\,M_\odot$ star rotating at 98\% of the critical angular velocity.  Here $p\_{\mathsf{gamma}}$ indicates the polar profile of $\gamma$, while $e\_{\mathsf{gamma}}$ indicates the equatorial profile. Values close to 0.4 indicate that convection is efficient (e.g., in the core). The decrease in $\nabla_{\mathsf{ad}}$ toward the surface stems from the partial ionization of helium, creating thin radial layers that are unstable to convection.
      \label{fig:del_ad}}
\end{minipage}%
\end{figure}

In Figure \ref{fig:del_ad} we illustrate the adiabatic gradient, $\nabla_{\mathsf{ad}}$, and an estimate of this gradient using the adiabatic index, $1 - 1 / \gamma$, both calculated by ESTER, as a function of temperature. The profiles represent the internal structure of a $1.65\,M_\odot$ star rotating at 98\% of the critical angular velocity. In regions where the convection is efficient, meaning that little energy is lost to ``radiative leakage'', then the actual gradient within a convective region will be very close to the adiabatic gradient. This is the case in the core where values are all close to 0.4. The sharp drop in $\nabla_{\mathsf{ad}}$ toward the surface stems from the partial ionization of helium.  Lower values of $\nabla_{\mathsf{ad}}$ make convective instability easier to achieve with a given temperature gradient (i.e., enabling $\nabla > \nabla_{\mathsf{ad}}$). While these regions appear broad when plotted versus logarithmic temperature (as in Figure \ref{fig:del_ad}), they are comparatively narrow in radial extent.

\subsection{The Entropy Gradient \label{subsentropy}}

\begin{figure}[h]
  \resizebox{6in}{!}{\includegraphics{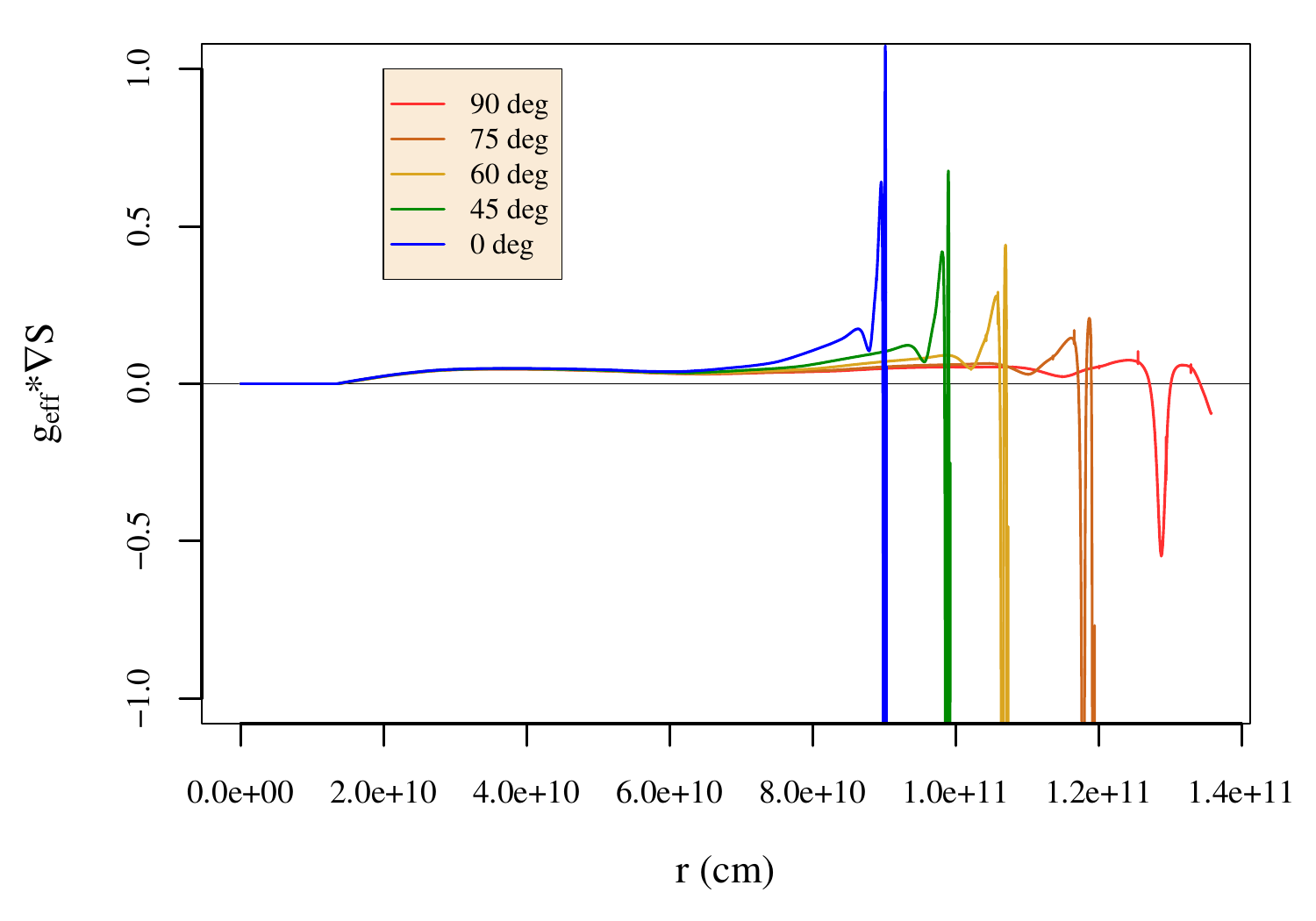}}
      \caption{Profiles of $g_{\mathsf{eff}} \cdot \nabla S$ for a $1.4\,M_\odot$ star rotating at 98\% of the critical angular velocity.  Regions where $g_{\mathsf{eff}} \cdot \nabla S < 0$ are convectively unstable. Values of zero near the center of the star confirm that the convection is efficient. All profiles have convectively unstable regions near the surface. Because of the cooler surface temperatures caused by gravity darkening, the equatorial profile ($\theta = 90 \degree$) exhibits a convectively unstable region that is both deeper below the surface and broader in radial extent than the  profiles at different $\theta$ angles.
 \label{fig:entropy_gradient}
      }
\end{figure}

An alternative criterion used to define regions of convective instability relies on determining whether the local entropy, $S$, increases or decreases outward. From thermodynamic principles it can be shown \citep[e.g.,][]{hansen2004} that if entropy decreases outward, that is if $\vec{g}_{\mathsf{eff}}\cdot\nabla S > 0$, then a region is convectively unstable.  And conversely, if $\vec{g}_{\mathsf{eff}}\cdot\nabla S < 0$, then the region is stable against convection.  \citet[][Section 2.1.3.]{rieutord2016} specifically state this criterion as 
\begin{eqnarray}\label{eqconvcrit}
- g_{\mathsf{eff}} \cdot \nabla S > 0~.
\end{eqnarray}
Since $g_{\mathsf{eff}}$ is always positive (Figure \ref{fig:geff_radius}), the negative sign flips the conditional inequality. In the special case of  efficient adiabatic convection, $\nabla S$ will be effectively zero through the convective zone.

Profiles of $\vec g_{\mathsf{eff}} \cdot \nabla S$ as calculated by ESTER are illustrated in  Figure~\ref{fig:entropy_gradient}. The figure shows this criterion vs radius for a $1.4\,M_\odot$ star rotating at 98\% of the critical angular velocity. Values of $g_{\mathsf{eff}} \cdot \nabla S$ are very close to zero near the center (at all values of $\theta$), indicating that efficient convection occurs in the convective core. Profiles at all values of the polar angle $\theta$ ($0^\circ$, $45^\circ$, $60^\circ$, $75^\circ$ and $90^\circ$) have a formally defined convectively unstable region where $\vec{g}_{\mathsf{eff}} \cdot\nabla S < 0$ near the outer surface; these profiles extend to different radii because the star's shape is rotationally distorted.  With the exception of the equatorial profile at $\theta = 90\degree$ (and possibly $\theta = 75\degree$), the convectively unstable region is thin and immediately below the surface of the star. 
Generally this is seen as too close to the surface for any significant energy transport via convection. For the equatorial profile at 90$\degree$, however, the temperature at the surface is much cooler because of gravity darkening. This results in a convectively unstable region at the equator that is well below the surface and broader in radial extent. \citet{espinosa_lara_rieutord2013} noted a similar trend. {Our stellar structure models calculated with ESTER suggest that for rapidly rotating intermediate mass stars with a very thin convective envelope at polar latitudes, rotation may cool their equatorial regions sufficiently to result in a convective envelope that is considerably thicker there, and which could sustain a dynamo. Modeling this more accurately will require advances in 2D models of convection for stellar structure calculations.}

\subsection{Brunt–V\"ais\"al\"a Frequencies}

A third criterion for convective instability can be determined directly from the net force acting on displaced parcel of stellar plasma. If the region is stable against convection, a restoring force will cause the parcel to oscillate about its original position.  The frequency of these oscillations is called the Brunt–V\"ais\"al\"a (BV) frequency
\citep{christensen2003stellar}. If the region is convectively unstable, a buoyancy force will cause the parcel to continue to move in the same direction (rising or sinking).  In this case the BV frequency is imaginary.  Commonly we assess convective instability by considering \textit{squared} BV frequencies, called $N^2$. In non-rotating stars this quantity is
\begin{eqnarray}
N^2=g\left[
{\frac {1}{\gamma }}{\frac {d\ln P}{dr}}-{\frac {d\ln \rho }{dr}}
\right] ~.
\end{eqnarray}
Here $g$ is the local gravity {(the modulus of gravity, thus positive)}, $P$ is pressure and $\rho$ is density. {This expression can be related to the entropy gradient, which is more convenient to manipulate, namely}
\begin{eqnarray}
 \nabla s = -c_p\left(\frac{\partial\ln
T}{\partial\ln\rho}\right)_{\! P}\frac{N^2}{g}\vec{e}_r~.
\end{eqnarray}
{This expression can be used to get an expression of the BV frequency that is also valid in a 2D model. We note}
\begin{eqnarray} 
\vec{g} \cdot \nabla s &=& c_p\left(\frac{\partial \ln  T}{\partial\ln\rho}\right)_{P}N^2 
\\
&=& -\frac{c_pN^2}{\delta} ~,
\end{eqnarray}
{where $\delta=\left(\frac{\partial\ln \rho}{\partial\ln T}\right)_{P}$.}
{The BV frequency is then}
\begin{eqnarray} 
N^2 = -\frac{\delta}{c_p} \vec g_{\rm eff}\cdot\nabla s~.
\end{eqnarray}

If $N^2 > 0$, the BV frequency is a real number and the region is stable against convection. Conversely, if $N^{2} < 0$, the BV frequency is an imaginary number and the region is convectively unstable.
\begin{figure}[h]
\begin{minipage}[c]{0.65\linewidth}
\includegraphics[width=\linewidth]{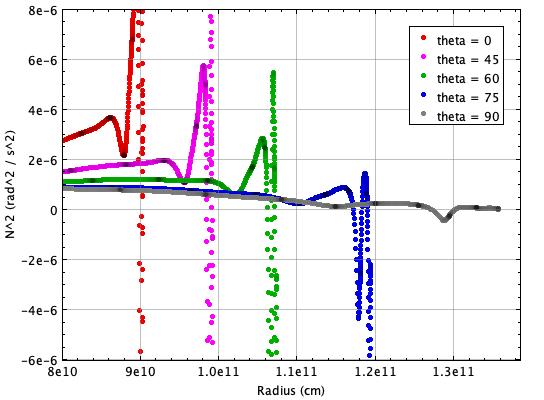}
\end{minipage}
\begin{minipage}[c]{0.35\linewidth}
      \caption{Profiles of squared Brunt-V{\"a}is\"al\"a (BV) frequency, $N^2$, versus radius for a $1.4\,M_\odot$ star rotating at 98\% of the critical angular velocity.  Regions where $N^2 < 0$ are convectively unstable.  A convectively unstable region develops below the surface at $\theta = 75 \degree$, and then becomes deeper (further from the surface) and broader (wider in radial extent) at the equator ($\theta = 90 \degree$).
      \label{fig:figN2_1p4}}
\end{minipage}
\end{figure}

Squared BV frequencies are illustrated in Figures~\ref{fig:figN2_1p4} in the outer portion of a $1.4\,M_\odot$ star rotating at 98\% of the critical angular velocity. These profiles demonstrate a trend in convective instability that is consistent with the entropy gradient profiles (Section \ref{subsentropy}). All of these radial profiles have a convectively unstable region near the surface. However, profiles closer to the equator have a convectively unstable region that is deeper (further from the surface) and broader in radial extent \cite[see][Figs 8,9]{espinosa_lara_rieutord2013}.

\begin{figure}[h]
\includegraphics[width=.5\linewidth]{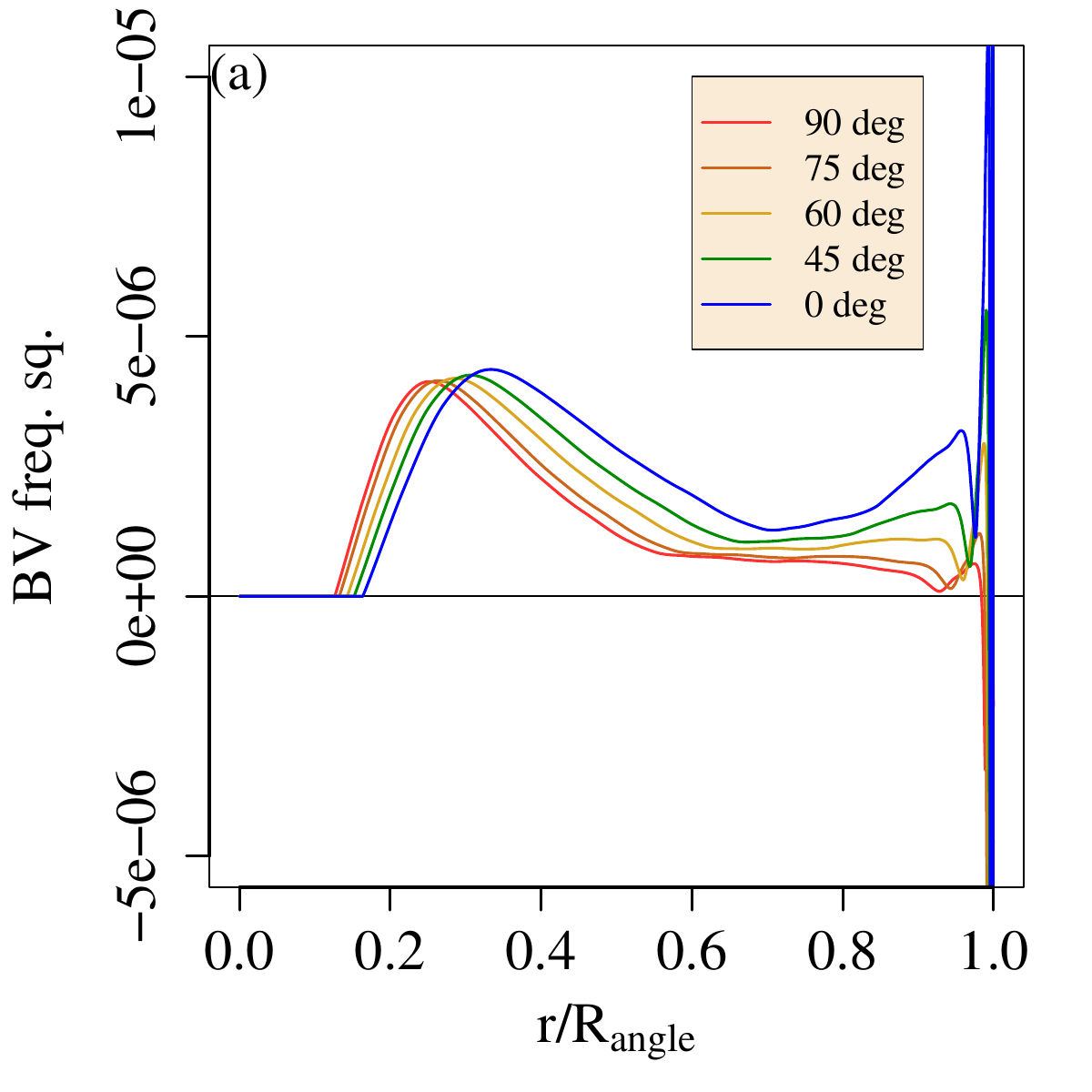}
\includegraphics[width=.5\linewidth]{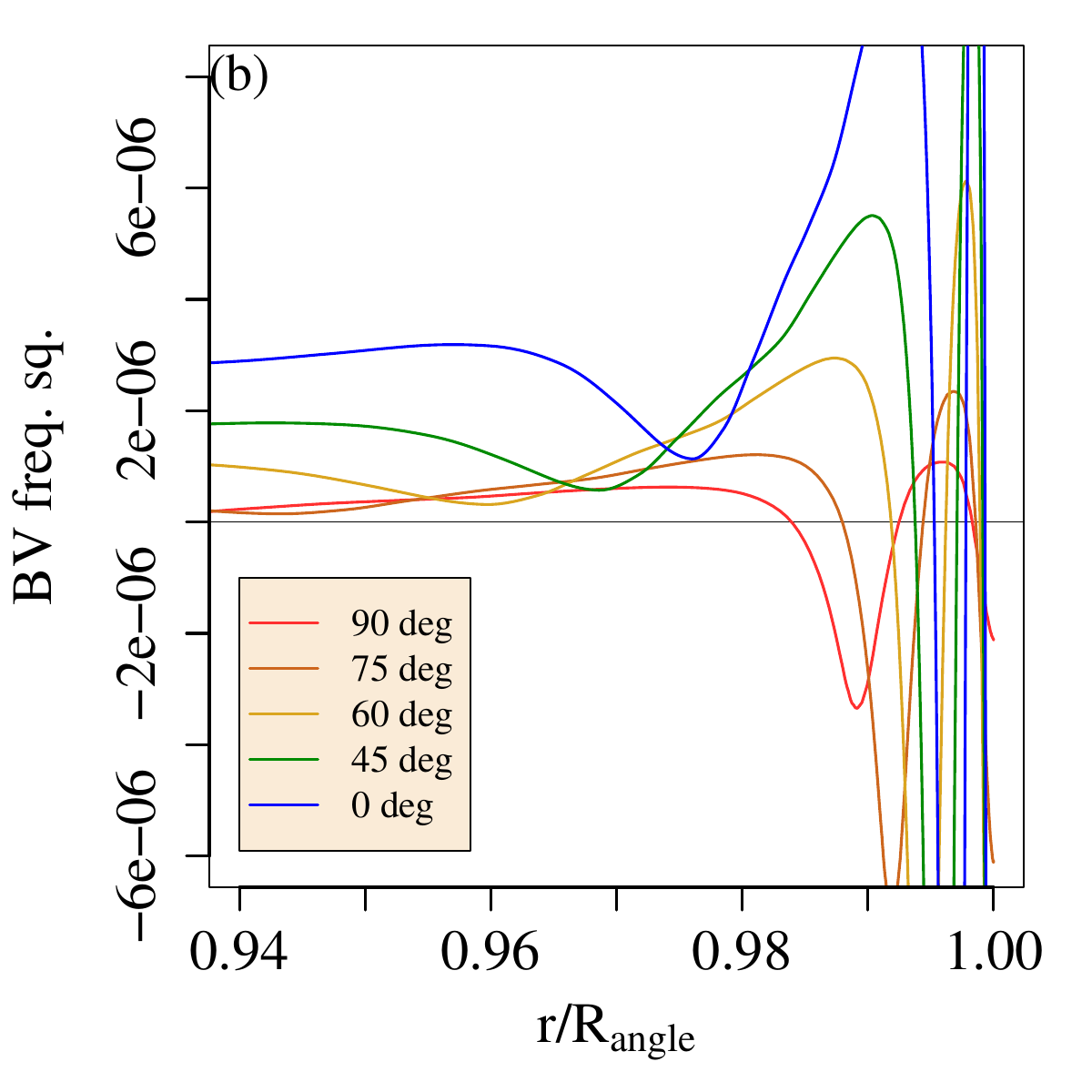}
      \caption{Profiles of squared Brunt-V{\"a}is\"al\"a (BV) frequency, $N^2$, versus normalized radius in the outer regions of a $1.86\,M_\odot$ star rotating at 80\% of the critical angular velocity; radial profiles at five different $\theta$ values are shown, ranging from polar ($\theta = 0 \degree$) to equatorial ($\theta = 90 \degree$).  In (a) it is clear that efficient convection occurs in the core ($N^2 \sim 0$), and the interior that is mostly radiative ($N^2 > 0$), with the exception of thin formally convective regions near the surface.  A zoomed in view near the surface  is shown in (b).  The equatorial profile has a convective region that is deeper (further from the surface) and broader (larger in radial extent) than the other profiles.
      \label{fig:N2_normalized_radius}}
\end{figure}

Squared BV frequencies are also illustrated in \ref{fig:N2_normalized_radius} versus normalized radius, in this case for a $1.86\,M_\odot$ star rotating at 95\% of the critical angular velocity. The squared BV frequencies are 
essentially zero near the center, confirming efficient convection in the convective core. For much of the remaining radial extent of the star, the squared Brunt-V{\"a}is\"al\"a frequencies are positive; the fluid is stable against convection. Small, formally convectively unstable regions appear close to the surface as helium and then hydrogen transition to being partially ionized. Toward the equator the convectively unstable region deepens (is further from the surface) and becomes broader (has a larger radial extent).  Detailed modeling of Cepheids with similar surface temperatures have shown that in addition to the thin sub-surface convective envelope seen here, some stars can develop a convectively unstable shell below the envelope (see \citet{stuck2025convective}).
\begin{figure}[h]
\begin{minipage}[c]{0.6\linewidth}
\includegraphics[width=\linewidth]{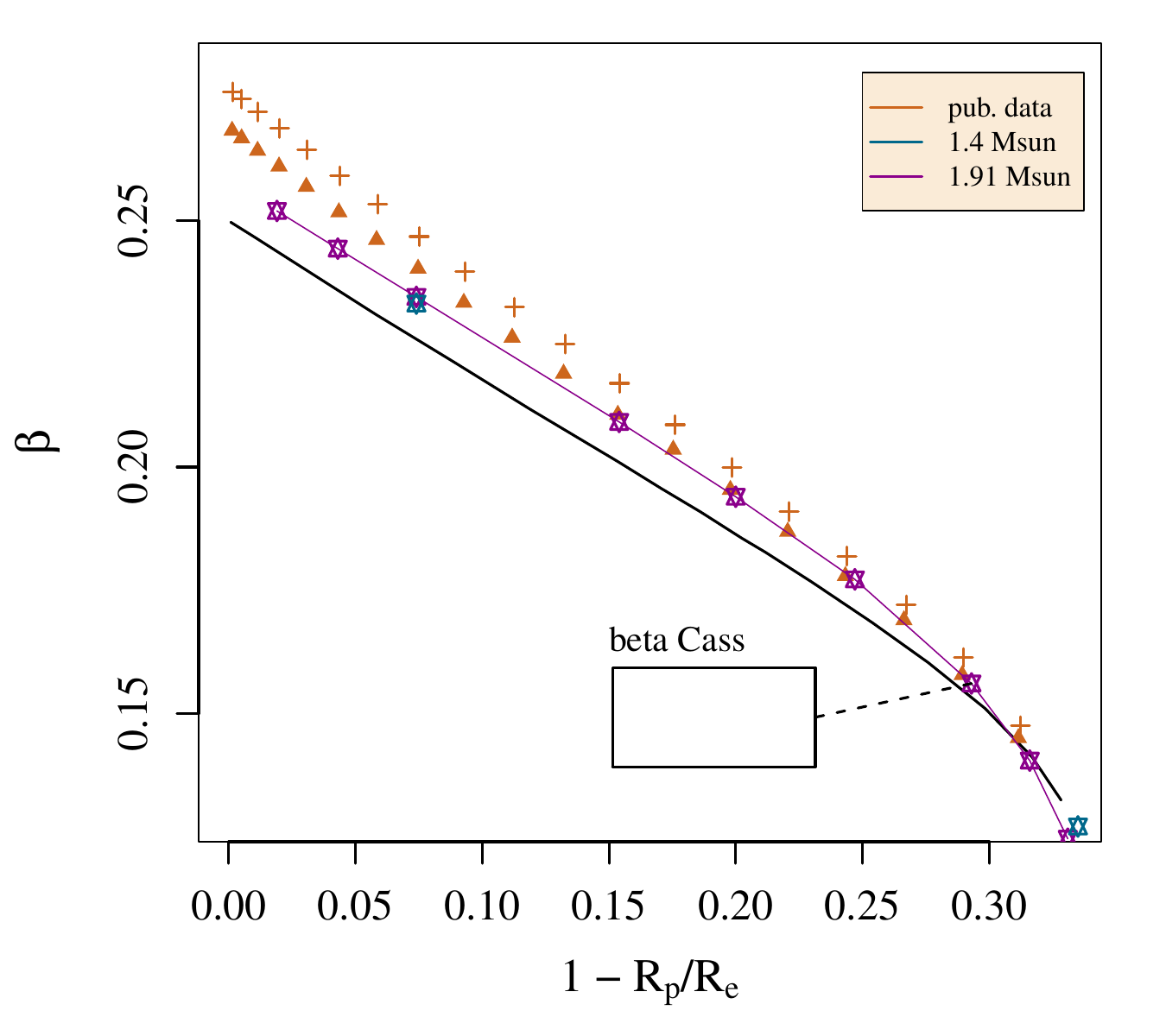}
\end{minipage}
\begin{minipage}[c]{0.35\linewidth}
      \caption{Gravity darkening exponent, $\beta$, as a function of rotational flattening. ESTER models of a $1.91\,M_\odot$ star are shown with $X_{\mathsf{c}} = X$ (\textit{orange crosses}), $X_{\mathsf{c}} = 0.5X$ (\textit{orange triangles}) and $X_{\mathsf{c}} = 3.57 X$ (\textit{purple stars}).  Both data sets in orange are  from  \citet{espinosa_lara2011gravity}.  A similar composition ($X_{\mathsf{c}} = 3.57 X$) and lower mass $1.4\,M_\odot$ star is included for comparison (\textit{blue stars}). The gravity darkening model proposed by \citet{espinosa_lara2011gravity} is shown as a \textit{black line}. None of these models match the value of $\beta$ measured for the cool rapid rotator $\beta$ Cassiopeiae \citep{che2011}, indicated by the \textit{open box}. A dashed line connects the measurement with an ESTER model rotating with 90\% of the critical angular velocity.
    \label{fig:gravitydarkening} }
\end{minipage}
\end{figure}

\section{ESTER gravity darkening predictions versus observations}

ESTER is able to reproduce the gravity darkened surface of a rapidly rotating star (see Figure~\ref{fig:esterviztemp}). This can be compared directly with images of rapidly rotating stars (see Figure \ref{fig:rapid_rotators}) to test theories of how effective temperature scales with local surface gravity (e.g., ${T}_{\mathsf{eff}}\propto {g_{\mathsf{eff}}}^{\beta}$). This in turn can constrain how energy is transported near the surface. \citet{espinosa_lara2011gravity} provide a comparison between the gravity darkened surfaces predicted by ESTER and those  measured via interferometric imaging for 4 rapidly rotating stars.  
We conduct a similar comparison of ESTER predictions with the star $\beta$ Cassiopeiae, imaged with the CHARA Array \citep{che2011}. With a spectral type F2IV, $\beta$ Cassiopeiae is the coolest rapidly rotating star imaged thus far. \citet{espinosa_lara2011gravity} estimate its mass to be $1.91\,M_\odot$. Its subgiant spectral classification suggests that it may be slightly evolved.

We can calculate the gravity darkened exponents, $\beta$, for a set of stars with a broad range of stellar angular velocities (see Figure \ref{fig:gravitydarkening}).  Each value of $\beta$ is calculated assuming the scaling law of von Zeipel (eq.~\eqref{eq:vonzeipel}), using ESTER values of ${T}_{\mathsf{eff}}$ and $g_{\mathsf{eff}}$ over a range of latitudes and by obtaining a least squares linear fit to 
\begin{eqnarray}
\ln({T}_{\mathsf{eff}})\propto \beta \ln{(g_{\mathsf{eff}})}~.
\end{eqnarray}
Here the angular velocity is quantified by how much the star is ``flattened'' from this rotation, as $1 - R_{\mathsf{p}}/R_{\mathsf{e}}$ (Section \ref{sec:rotationeffects}).

The $\beta$ values in Figure \ref{fig:gravitydarkening} are for 4 stars, with 3 of these having a mass of $1.91\,M_\odot$, the value adopted for $\beta$ Cassiopeiae.  These are calculated at different stages of evolution by specifying different core hydrogen fractions ($X_{\mathsf{c}} = 0.5X, X$ and $3.57 X$).  The fourth star has a lower mass of $1.4\,M_\odot$ and a 
core mass fraction of $X_{\mathsf{c}} = 3.57 X$.  All stars exhibit a similar trend of decreasing values of $\beta$ with increased rotation.  At low angular velocities ($1 - R_{\mathsf{p}}/R_{\mathsf{e}} < 0.10$), significant variations in the core hydrogen fraction change the gravity darkening exponent by a few percent. All models converge to a similar gravity darkening exponent at high angular velocities ($1 - R_{\mathsf{p}}/R_{\mathsf{e}} \sim 0.30$). Interestingly, both the $1.4\,M_\odot$ and 
and $1.91\,M_\odot$ ESTER models have similar gravity darkening exponents at all angular velocities.  At a given core hydrogen mass fraction, the gravity darkening laws produced by our ESTER calculations do not appear to be strongly mass dependent.  Figure \ref{fig:gravitydarkening} also shows the model proposed by \citet{espinosa_lara2011gravity} for comparison. This model disagrees with ESTER results by $\sim 10\%$ at low angular velocities but likewise converges at high angular velocities.

Perhaps most significant is that neither the ESTER models nor the model of \citet{espinosa_lara2011gravity} agree with the observed value of $\beta$ measured for the rapidly rotating star $\beta$ Cassiopeiae \citep[][Figure \ref{fig:gravitydarkening}]{che2011}. The value of $\beta$ calculated from observations is lower than what all models predict at that angular velocity. As noted above, $\beta$ Cassiopeiae is the coolest of the rapid rotating stars imaged thus far. With a surface temperature just warmer than the Kraft Break, it may have a convection zone with a thickness that is latitude dependent.  A more sophisticated physical model that incorporates convection may be needed to represent its structure and the resulting gravity darkening.

\section{\label{sec:bench} Benchmarking ESTER results against MESA}

\begin{figure}
\begin{center}
\includegraphics[width=.8\linewidth]{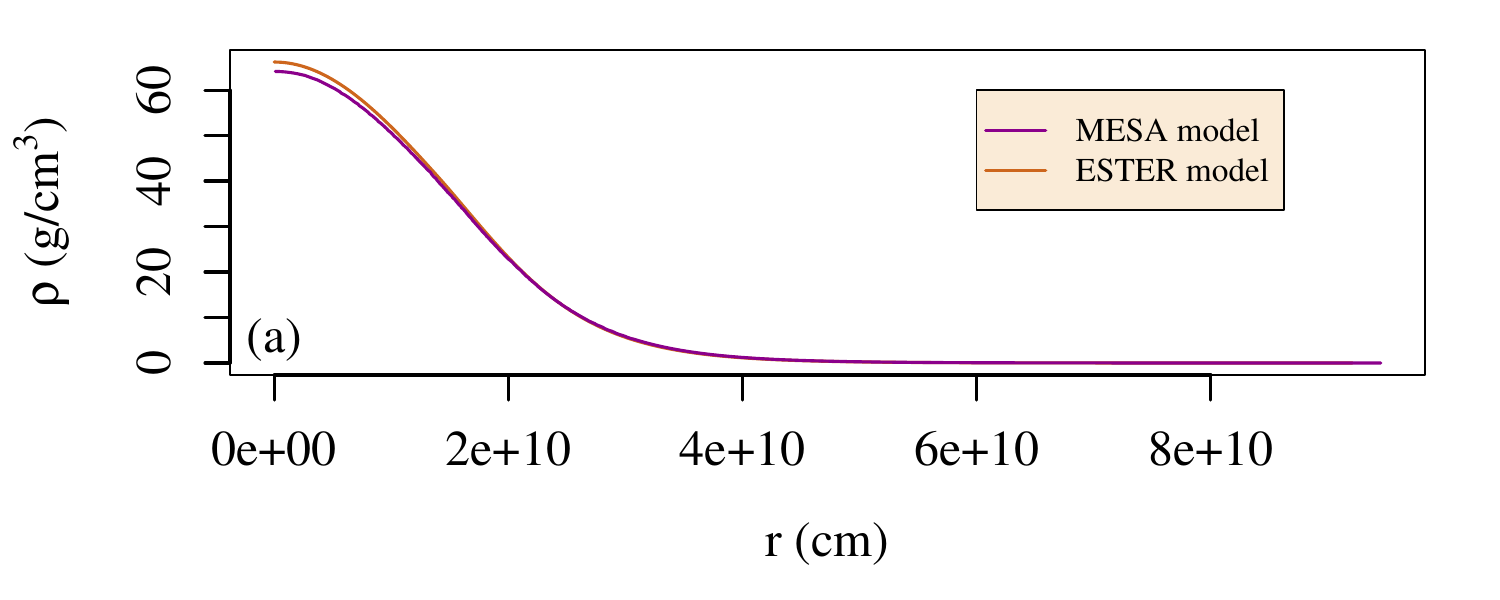}
\includegraphics[width=.8\linewidth]{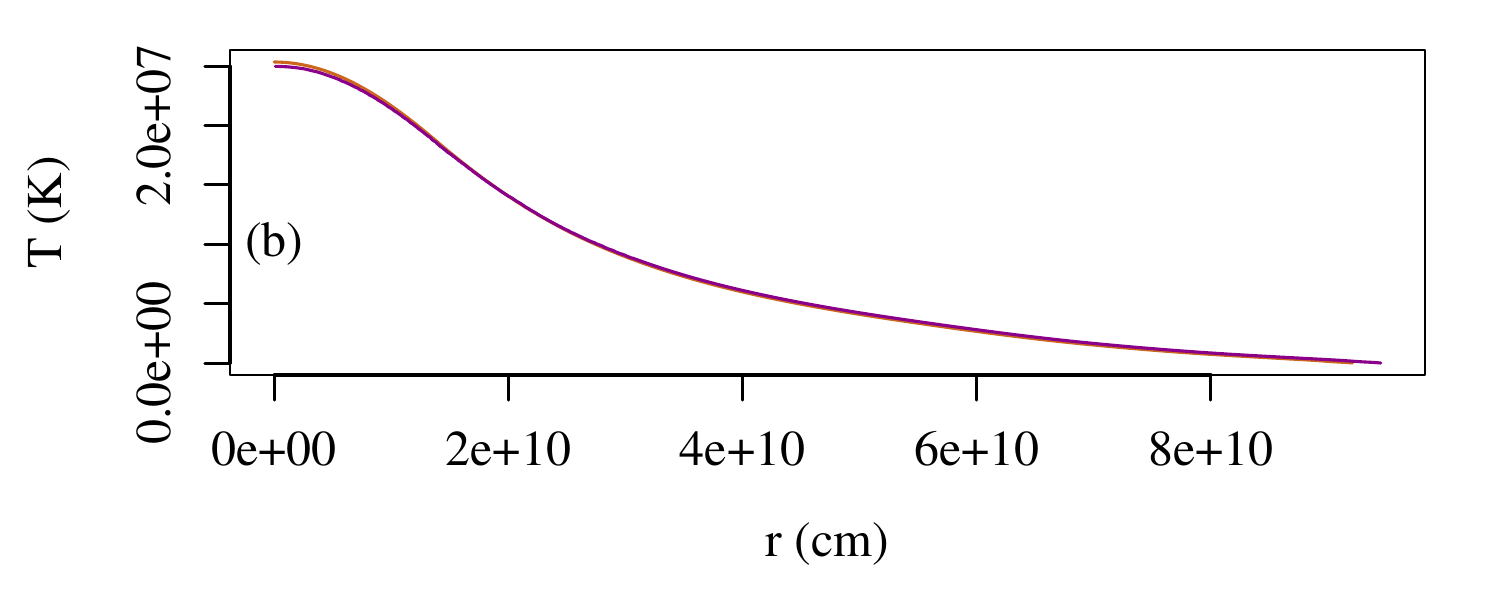}
\includegraphics[width=.8\linewidth]{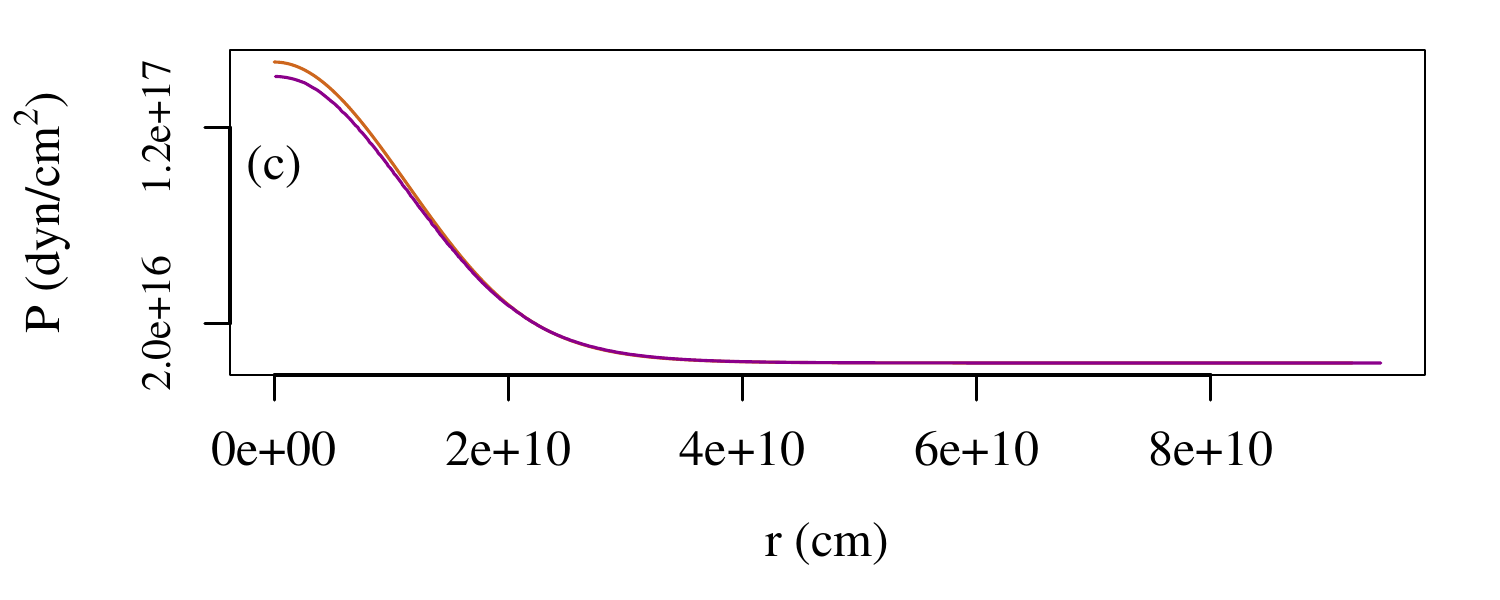}
\includegraphics[width=.8\linewidth]{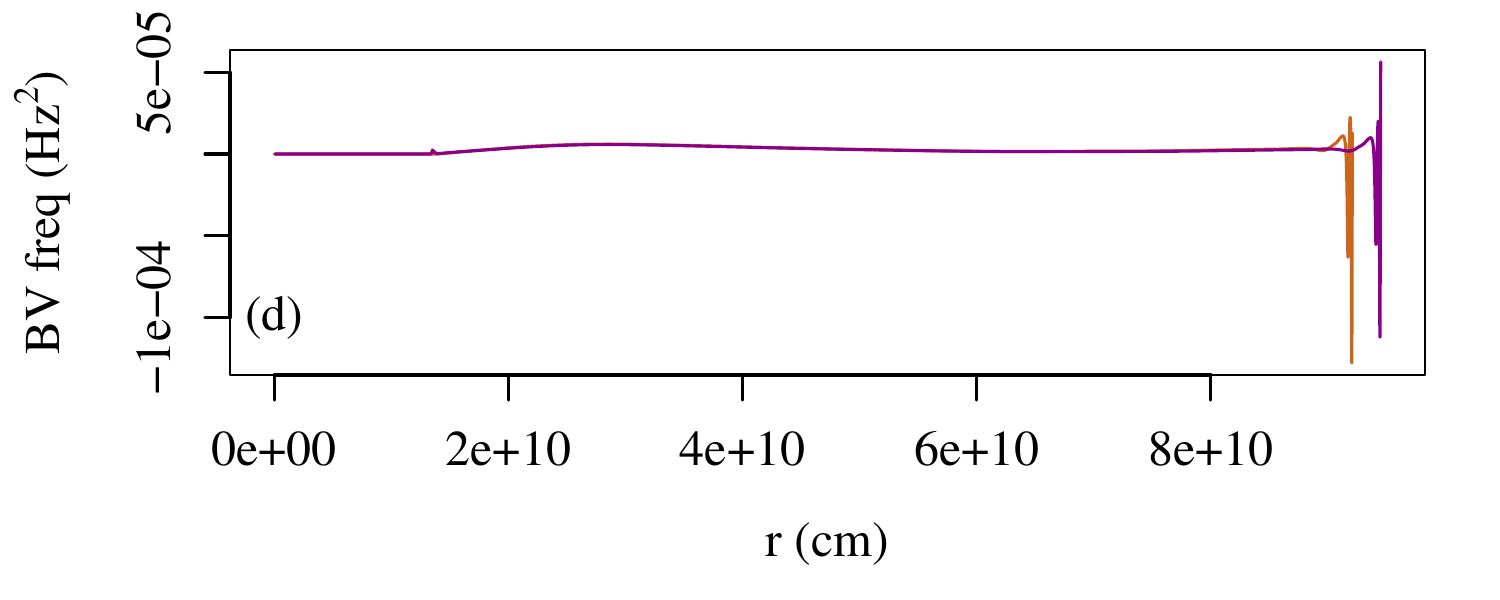}
      \caption{Stellar profiles of density, temperature, pressure and squared Brunt-V{\"a}is\"al\"a frequency for a $1.4\,M_\odot$ star computed by ESTER and MESA. Neither star is rotating and the adopted physics for each model is chosen to be as similar as possible and to match an evolutionary state at the zero age main sequence (see text for details). The similarity of these profiles is impressive given that the governing equations are formulated so differently.
      \label{fig:mesacomp1p4}}
\end{center}
\end{figure}
The 1D stellar structure and evolution code MESA \citep[Modules for Experiments in Stellar Astrophysics;][]{paxton2010modules} is used widely in stellar astrophysics because it is open source and it allows for the input physics to be tuned. Comparisons of MESA model predictions with observations of stars continue to improve its viability. It thus provides a useful benchmark to test ESTER's structure calculations. However, since the version of ESTER that we explore here only produces a stellar structure and does not model the time evolution of that structure, comparisons of ESTER models with MESA models must be done at a specific evolutionary state (e.g., age). This kind of direct comparison of stellar structure models has rarely been undertaken, but is extremely valuable \citep[e.g.,][]{aguirre2020aarhus}.

In Figure~\ref{fig:mesacomp1p4} we illustrate stellar profiles of density, temperature, pressure and squared Brunt-V{\"a}is\"al\"a (BV) frequencies of 
a $1.4\,M_\odot$ star as calculated by MESA and by ESTER. The MESA model corresponds to a star near the zero age main sequence (ZAMS) point on the MESA evolutionary track; it has an age of 3.54 million years, specifically. An ESTER model with no hydrogen depletion in the core (e.g., $X_{\mathsf{c}} = X$) is chosen to match this. Both use a ``simple'' atmosphere model, the OPAL equation of state and opacity tables with a solar mixture. No rotation is assumed in either model and both are calculated as 1D.

The stellar profiles are remarkably similar, with the ESTER model having only slightly higher values of density, temperature and pressure in the core than the MESA model. The squared BV frequency profile is likewise similar, indicating convective instability very near the surface. The ESTER stellar structure is 2.6\% smaller than the MESA stellar structure, which is less than the typical differences between stellar models and observations. Overall, the strong similarity between the stellar structures produced by these two codes is impressive, given that the governing equations are formulated so differently. A deeper investigation of the products of these codes would be necessary to understand the causes of the small differences we observe here.

\begin{figure}
\begin{minipage}[c]{0.65\linewidth}
\includegraphics[width=\linewidth]{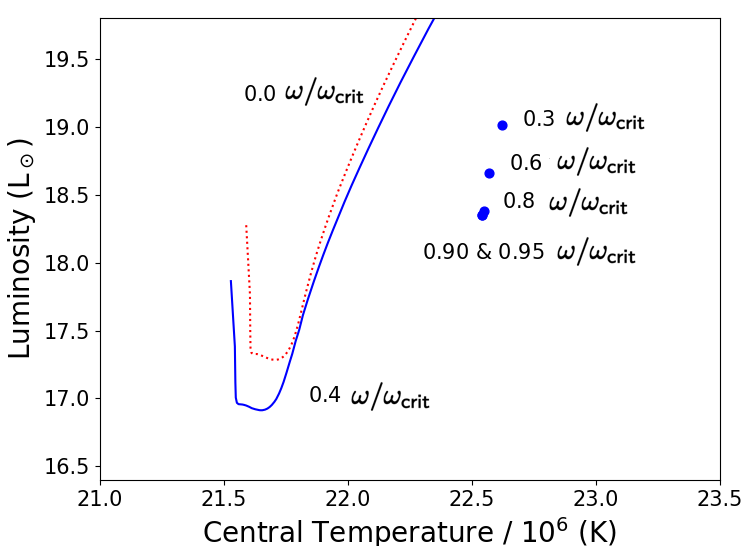}
\end{minipage}
\begin{minipage}[c]{0.35\linewidth}
      \caption{
      Values of luminosity (tracing energy generation) and central temperature from ESTER models of a $2\,M_\odot$ star at the ZAMS, calculated for 5 different angular rotation rates are shown as blue data points.  For comparison,  MESA evolutionary tracks starting at the ZAMS are shown for models with no rotation (\textit{dotted red line}) and with rotation at 40\% of the critical (\textit{solid blue line}). 
      \label{fig:Explore_MIST}}
\end{minipage}%
\end{figure}
We also examine how rotation of a $2\,M_\odot$ star affects its luminosity and the central values of temperature, density and pressure. Figure~\ref{fig:Explore_MIST} illustrates the total luminosity (tracing energy generation) and central temperature for 2 MESA models calculated at 0 \% and at 40\% of the critical angular velocity. For comparison, ESTER models at 30\%, 60\%, 80\%, 90\% and 95\% of the critical angular velocity are shown.  
The ESTER models are $\sim$ 5\% more luminous and $\sim$ 5\% hotter than the corresponding MESA models; both models decrease in luminosity and central temperature as rotation increases. Table \ref{tab:2msun_central} lists the total luminosity, core temperature, core density, and core pressure of $2\,M_\odot$ ESTER models rotating up to 0.95 of the critical angular velocity. The luminosity decreases by 3.4\% as the rotation increases from 30\% to 90 - 95\% of the critical angular velocity.
\begin{table}
\rowcolors{2}{gray!10}{white}
\centering
\caption{Luminosities and central values for rotating $2\,M_\odot$ stars \label{tab:2msun_central}}
\begin{tabular}{ccccc}
             & Luminosity  & $T_{\mathsf{c}}$      & $\rho_{\mathsf{c}}$   & $P_{\mathsf{c}}$ \\
$\omega / \omega_{\mathsf{cr}}$ & ($L_\odot$) & ($10^7$ K) & (g/cm$^2$) & ($10^{17}$ dyn/cm$^2$) \\
\hline
0.00         & 18.96       & 2.265      & 74.42      & 1.902 \\
0.30         & 19.01       & 2.262      & 74.74      & 1.908 \\
0.60         & 18.66       & 2.257      & 75.34      & 1.919 \\
0.80         & 18.38       & 2.255      & 75.61      & 1.924 \\
0.90         & 18.35       & 2.254      & 75.67      & 1.925 \\
0.95         & 18.35       & 2.254      & 75.69      & 1.925 \\
\hline
\end{tabular}
\end{table}

\section{\label{sec:end} Summary and Discussion}

This technical report has been formulated with the goal of helping students and researchers get started using the 2D stellar structure code ESTER.  We have described the equations that ESTER solves, the numerical method it uses, and the kind of output that our local version of the code produces.  We have compared ESTER's steady-flow stellar structure equations to standard one-dimensional stellar structure and evolution codes.  We have produced two-dimensional ESTER models that exhibit gravity darkening when rapidly rotating. These same models indicate that a region of convective instability could exist in the outer envelope of some stars, and that the convection zone becomes deeper and broader toward the equator. Zones of convection that are substantial only at the equator are plausible.  As discussed in \citet{espinosa_lara2011gravity}, ESTER is not able to reproduce the the gravity darkening exponent measured for the star $\beta$ Cassiopeiae; we suggest that $\beta$ Cassiopeiae, the coolest rapid rotator interferometrically imaged thus far, may sustain {a sub-surface convective layer of significant depth near the equator}.  This implies that there are a subset of stars that are rapidly rotating, for which a model for {envelope} convection in ESTER would be needed.

 By comparing one-dimensional non-rotating ESTER stellar structures to MESA stellar structures, we have demonstrated that ESTER produces stellar structures that are sound. In most cases differences between the models are smaller than current differences between models and observations; it would not yet be possible to tell which model is ``best''. The general agreement between the models, despite their very different formulation is encouraging both for the use of 1D stellar models and the further development of 2D stellar models. We have also shown how ESTER and MESA stellar structures change as the angular velocity of the model is increased.  At this time, ESTER does not contain the wide range of different physical models implemented in MESA. However for many stars, direct comparisons of structures from these codes show the predictions to be similar.

\appendix 
\section{\label{sec:appA} Appendix A: Compiling ESTER}

The wiki associated with the ESTER project \href{https://github.com/ester-project/ester/wiki/Install} has several descriptions
for how to compile or ``install'' the code.  Depending on where you are trying to compile this code,
one method may work more readily.  On all of the Livermore computing (LC) machines we have tried, the cmake build has worked straightforwardly.  We refer users on LC to the file in our local repository named TOCOMPILE.

The following description is for compiling the code on the Harlow cluster at Georgian State University using the autoconfig method to build the code.  ESTER uses several commonly used libraries: BLAS, CBLAS and LAPACK.  These are also included in the intel-mkl library.  On harlow, you will need to load the following modules to load these libraries:  1) intel/18.0.3.222   2) fftw/3.3.8   3) openfft/1.2   4) hdf5/1.10.1  .  The command to load these is
\begin{verbatim}
module load intel/18.0.3.222
module load fftw/3.3.8
module load openfft/1.2
module load hdf5/1.10.1 
\end{verbatim}
We then pursue the installation that uses autoconfig.  First you create a directory for the build:
\begin{verbatim}
mkdir local
\end{verbatim}
Then, within the main trunk of the code, run bootstrap.
\begin{verbatim}
./bootstrap
\end{verbatim}
Then replace your current directory in the prefix flag below and execute the configure command
\begin{verbatim}
CC=mpiicc FC=mpiifort FCFLAGS="-O3 -fPIC" CXXFLAGS="-O3"  ./configure
--with-hdf5="/opt/ohpc/pub/libs/intel/hdf5/1.10.1/bin" 
--prefix=/data/jpratt/testpester/nnew/pester/local 
LIBS="-L/opt/intel/compilers_and_libraries_2018.3.222/linux/mkl/lib/intel64/
-lmkl_blas95_lp64 -lmkl_lapack95\_lp64
-L/opt/ohpc/pub/libs/intel/hdf5/1.12.0/lib/
-I/opt/ohpc/pub/libs/intel/hdf5/1.12.0/include/ 
-I/opt/intel/compilers_and_libraries_2018.3.222/linux/mkl/include/ 
-I/opt/intel/compilers_and_libraries_2018.3.222/linux/mpi/include64/mpi.h"
\end{verbatim}

\begin{verbatim}
make -j8
make install
\end{verbatim}
This configures the compilation so that it will link to the correct libraries and use the correct (Intel) compilers.  When this configure exits successfully, you can compile the code.  I suggest that you use 8 processors because the compilation is moderately long
\begin{verbatim}
make -j8
\end{verbatim}
When this compilation is successful (it does not end with an error message), there is still no executable to run.  To produce the executable you must do
\begin{verbatim}
make install
\end{verbatim}
This will put an executable in the directory that you specified in the prefix, which is ``local'' in this example.

\section{\label{sec:appB} Appendix B: Input Parameters}

To run ESTER, one needs two different parameter input files (analogous to the MESA inlist).  One of these is to produce a 1D stellar structure, and the other is produce a 2D stellar structure, which requires the output from an ESTER stellar structure as input.  Note that ESTER is c++, where MESA is f90; MESA inlists are collections of fortran namelists, and the namelist structure for input parameters is not available in c++.  These files are thus not the same structure as an inlist.  Here below we provide a 1D set-up with our comments

\begin{verbatim}
ndomains=15 # the number of sub-domains to be used
npts=60  # the number of points in each sub-domain

M=1.4          # mass of star
X=0.28         # fraction of hydrogen
Z=0.02         # Mass fraction of metals

Xc=1           #hydrogen fraction in convective core
surff=1        #truncate simulation at point below surface

Tc=1.901477e7  #initial central temperature (units)
pc=2.196128e17 #initial central pressure

opa=opal       # OPAL opacity tables.  
# Other choices for opacity are: houdek, kramer
eos=opal       # equation of state that corresponds with OPAL tables.
#Other choices for eos are: ideal, ideal+rad
nuc=simple     # nuclear reactions.  Other choices: cesam
atm=simple     # model of star's atmosphere

core_convec=1  # Deactivate core convection with core_convec=0
# Minimum size of convective core (in fraction of polar radius)
min_core_size=0.01
env_convec=0

init_poly=0
\end{verbatim}

Here below we provide a 2D parameter file set-up with rotation
\begin{verbatim}
ndomains=15 # the number of sub-domains to be used
npts=60  # the number of points in each sub-domain

nth=24  # number of grid points in latitude (theta)
nex=30  # Number of radial points in the external domain

M=1.4          # mass of star
X=0.28         # fraction of hydrogen
Z=0.02         # Mass fraction of metals

Xc=1           #hydrogen fraction in convective core
surff=1        #truncate simulation at point below surface

Tc=2.523600e7   #initial central temperature (units)
pc=1.555767e17  #initial central pressure

stratified_comp=1
surff=1  #truncate simulation at point below surface

opa=opal    # OPAL opacity tables.  
# Other opacity choices: houdek, kramer
eos=opal    # equation of state that corresponds with OPAL tables.
# Other EOS choices: ideal, ideal+rad
nuc=simple  # nuclear reactions.  Other choices: cesam
atm=simple  # model of star's atmosphere

# Angular velocity at the equator in units of the critical velocity
Omega_bk=.98 

#  Ekman number: dimensionless number used in fluid dynamics
# to describe the ratio of viscous forces to Coriolis forces
# In ESTER this gives the amplitude of the meridional circulation.
Ekman=1  

core_convec=1        # Deactivate core convection with core_convec=0
# Minimum size of convective core (in fraction of polar radius)
min_core_size=0.01 
env_convec=0

init_poly=0
\end{verbatim}

\section{\label{sec:appC} Appendix C: Running ESTER}

Generally the procedure for creating a 2D ESTER model is
\begin{enumerate}
\item Run ESTER in 1D mode.
\item Run ESTER in 2D mode with 1D model as input.
\item Run ESTER in 2D mode with 2D model as input, but change the parameters slightly, i.e. by increasing
the angular velocity, or changing the mass fractions.
\item Repeat Step 3 until you have a model for the star you are happy with, or until the code fails to converge.
\end{enumerate}
Note that ESTER uses an iterative method to find an optimal structure and shape for a given star.  That iterative method sometimes fails to find any solution that satisfies the input parameters.

As a final step, once a satisfactory 2D model is calculated, there is a program that converts the native ESTER output format to a vtk file (\url{https://vtk.org/}) which can be visualized with the visualization software VisIt (\url{https://sd.llnl.gov/simulation/computer-codes/visit}) which is produced at LLNL and universally installed on Livermore Computing clusters.  An alternative visualization package, ParaView (\url{https://www.paraview.org/}) also works to visualize these kinds of files.

\section{\label{sec:numerics} An astronomer-friendly discussion of numerical methods}

The numerical methods used in ESTER are described in detail in \citet{rieutord2016algorithm}, and summarized in Section 2.5 of
\citet{espinosa_lara_rieutord2013}.  The main features of ESTER's numerical methods are:
\begin{itemize}
\item The grid is split into spheroidal shells so that the pressure ratio between the upper and lower
boundaries of each domain is approximately equal for all domains.  The spatial grid includes some number of shells interior to the star (specified by the user), as well as a domain external to the star.  
\item ESTER uses a spectral element method.  The most famous spectral element code is the open-source code Nek5000 \citep{fischer2007nek5000,saha2021review} and its more recent version NekRS \citep{fischer2022nekrs}.  Spectral element methods are a combination of finite element methods and spectral methods.
\item Any spectral method requires a fast Fourier transforms (FFT) library. On the Harlow cluster, ESTER is built with the MKL libraries which include a FFT library.  The open-source FFTW library is also a possibility.
\item ESTER uses an iterative method, the Newton-Raphson method, to solve the matrix equations that result from the spectral element method.  A simple preconditioner -- a matrix that multiplies the whole equation -- is also used to make the matrix inversion more efficient.
\end{itemize}

\section*{Acknowledgements \label{sec:ack}}
{\small  The authors would like to thank the LLNL Faculty Mini-Sabbatical Program, which supported R.W.'s time while studying this software and producing this educational report.
This work was performed under the auspices of the U.S. Department of Energy by Lawrence Livermore National Laboratory under Contract DE-AC52-07NA27344.  LLNL-TR-2007815. \\
\noindent M.R. acknowledges support from the European Research Council (ERC)
under the Horizon Europe program (Synergy Grant agreement N◦101071505:
4D-STAR). While partially funded by the European Union, views and
opinions expressed are however those of the authors only and do not necessarily
reflect those of the European Union or the European Research Council. Neither
the European Union nor the granting authority can be held responsible for them. 
}

\bibliographystyle{unsrtnat}
\bibliography{ester}

\end{document}